\documentclass[preprint,review,12pt]{elsarticle}

\usepackage{amssymb}
\usepackage{amsmath}

\usepackage{lineno} 
\usepackage{tikz} 
\usepackage{multicol} 
\usepackage{subcaption}  
\usepackage{url}  

\journal{Nuclear Inst. and Methods in Physics Research, A} 

\definecolor{darkgreen}{HTML}{008800}

\begin{document}

\begin{frontmatter}



\title{Treatment of material radioassay measurements in projecting sensitivity for low-background experiments} 



\author[pnnl]{R.H.M.~Tsang \corref{cor1}}
\author[pnnl]{I.J.~Arnquist}
\author[pnnl]{E.W.~Hoppe}
\author[pnnl]{J.L.~Orrell} 
\author[pnnl]{R.~Saldanha}

\cortext[cor1]{Corresponding author, heiman.tsang@pnnl.gov}
\address[pnnl]{Pacific Northwest National Laboratory, Richland, WA, USA}

\begin{abstract}
%
By analyzing sensitivity projections as a statisical estimation problem,
we evaluated different ways of treating radioassay measurement results (values and upper limits)
when projecting sensitivity for low-background experiments.
We developed a figure of merit that incorporates a notion of conservativeness to 
quantitatively explore the consequences of attempts to bias sensitivity projections,
and proposed a method to report sensitivity.

\end{abstract}

\begin{keyword}
low background experiments \sep material assay measurements \sep upper limits and uncertainty

\PACS 95.35.+d \sep 23.40.-s \sep 29.40.-n \sep 02.50.-r \sep 06.20.Dk

\MSC[2010] 62K99 \sep 62P35 \sep 65G99 \sep 65Z05

\end{keyword}

\end{frontmatter}


\renewcommand{\arraystretch}{1.25}

\section{Introduction}
\label{sec:Introduction}
Modern low-background experiments in dark matter and neutrinoless double beta decay research 
rely on material assay measurements of trace level radioactive impurity concentrations 
to model and project their anticipated experimental sensitivity. 

Experimental collaborations have reported detailed records of their material assay measurement programs 
(e.g., \cite{ABGRALL201622,NextPrior,PandaxPrior,CuorePrior, Cuore0Prior,NexoPrior, LEONARD2008490, Xenon100}) 
in preparation for construction of their experiments. 
A research community online database \cite{LOACH20166} 
has collected these and other measurements to provide a resource for searching and identifying 
material assay measurements across multiple low background experiments for a wide variety of materials. 
Researchers use these references to 
estimate potential levels of trace radioactivity that 
may be present in experimental conceptualizations. 

In the ideal scenario, experimental sensitivity projections would use reported material assay measurement \textit{values} 
well above the methodological detection limit of the assay measurement apparatus, 
with respect to their quantified \textit{uncertainties}.
However, in many cases experimental collaborations are finding and selecting ultra-pure materials 
that can only be quantified as having a radioimpurity concentration level given as an \textit{upper limit}, 
dependent upon the material assay methodological detection limit. 
As a result, next generation low-background experiments in dark matter and neutrinoless double beta decay research
produce sensitivity projections using a mixture of material assay measurement \textit{values} and \textit{upper limits}. 

Some researchers choose to be ``conservative'' and set the upper limit as a \textit{value} while
others choose to transform the upper limit to a distribution, 
either uniform or some variations of the Gaussian distribution, 
effectively treating the radioimpurity concentration like a Bayesian prior.
This adds some Bayesian elements into an otherwise completely frequentist procedure.
Naturally one expects these different choices to result in different predictions of experimental performance. 
To our knowledge, the impact of these choices has not been studied in a systematic fashion.

In the following sections, we describe our study on the impact of these choices.
We begin by illustrating the problem using a simple example in Section \ref{sec:SimpleExample}. 
Then, in Section \ref{subsec:AssayMethods},
we briefly review two common assay techniques and discuss their similarities and differences.
Next, in Section \ref{sec:Model}, we describe a Monte Carlo model of a typical assay campaign 
from material selection to sensitivity projection, and the different ways to treat assay values and upper limits.
The results of the Monte Carlo model are shown in Section \ref{sec:Scenarios}.
In Section \ref{sec:Discussion}, 
we define a figure of merit to evaluate those choices,
apply it to various scenarios,
and propose a method to report sensitivity.
Finally, we summarize in Section \ref{sec:Conclusion}.

\section{Combining a measurement with an upper limit}
\label{sec:SimpleExample}
To begin with a very simplified example, consider a small experimental arrangement composed of two materials, $A$ and $B$. Material $A$ was evaluated through a material assay method and was determined to contain a concentration of radioimpurity $X$ at a level of $11.1 \pm 1.2$~$\mu$Bq/kg. Material $B$ was evaluated in the same fashion and was determined to contain a concentration of radioimpurity $X$ at a level of \textit{less than} $0.8$~$\mu$Bq/kg, 
depending on the technology used and the institution reporting the limit.

From a high statistics radiation transport simulation, it is determined that materials $A$ and $B$ will contribute to the background event rate of the small experimental arrangement with approximately a ratio of 1:10.2,. This implies that both materials will contribute to the total background rate at roughly equivalent rates, if one presumes the actual \textit{true} radioimpurity level of material B is at or just below the instrument sensitivity. Table~\ref{tab:SimpleExperiment} captures the key inputs for projecting the background event rate in this small experimental arrangement.

\begin{table}[ht!]
\centering
\begin{tabular}{c c | c | c c c}
\multicolumn{6}{c}{} \\
\hline
Material & Mass & Radio-impurity $X$ & \multicolumn{3}{c}{Simulation information} \\ 
              & [kg]    & [$\mu$Bq/kg]       & Primaries & Background Events & Efficiency, $\varepsilon$ [Hz/Bq] \\ \hline
$A$        & 10      & $11.1 \pm 1.2$        & $10^{8}$  & 100,234 & 0.00100 \\
$B$        & 10      & $<0.8$                & $10^{8}$  & 1,022,387 & 0.01022 \\
\hline
\end{tabular}
\caption{\label{tab:SimpleExperiment}
Information from material assay measurements and high statistics radiation transport simulation for a small experimental arrangement composed of two materials, $A$ and $B$. This simplified experimental example is entirely hypothetical, deliberately constructed to elucidate the issue under study in this paper.}
\end{table}

The question arises: how to quantitatively combine the radioimpurity concentration of $X$ in materials $A$ and $B$ in projecting the expected experimental background rate in this small experimental arrangement? It is clear that, if we only consider material $A$, we should expect a background event rate, $R_A$, contribution of approximately $1.1\times10^{-7}$~Hz, or about $3.5\pm0.4$ background events for an uninterrupted year of data collection. These values, respectively, are determined using the following two equations:

\begin{equation}
R_i = \varepsilon_i \cdot M_i \cdot X_i 
\label{eq:SimpleRate}
\end{equation}
\begin{equation}
C_i = R_i \cdot \tau
\label{eq:SimpleCounts}
\end{equation}
where $i$ refers to material A or B; 
The variables
$\varepsilon_i$,
$M_i$, and
$X_i$,  
are 
the hit efficiency,
the mass, and
the radioimpurity concentration, 
of material $i$, with example values given in Table 1.
The quantity $\tau$ is the livetime of the experiment and 
$C_i$ is the number of background events expected in the experiment due to material $i$.

If we assume the worst case (or ``most conservative'' assumption) regarding the background event rate contribution from material $B$, we employ the same two equations (\ref{eq:SimpleRate}) and (\ref{eq:SimpleCounts}) and insert the \textit{upper limit} into the calculation. Doing so suggests a background event rate contribution of approximately $0.8\times10^{-7}$~Hz, or about $2.6$ background events for an uninterrupted year of data collection. 
Note it is now not clear 
how to represent the uncertainty on the number of background events in a year of data collection 
since we do not know the uncertainty 
associated with the underlying measurement of the radioimpurity concentration of $X$ in material $B$.

To devise a way to treat an upper limit, 
we first need to understand why an upper limit is reported and how it is calculated, 
and this depends on the assay technique used.
(For ease of discussion, all upper limits are at 90\% C.L. unless stated otherwise.)

\section{Assay methodologies}
\label{subsec:AssayMethods}

Many material assay techniques are employed by collaborations developing low background experiments 
as a means to measure ultra-trace levels of radioimpurities in materials
as an evaluation of feasibility in advance of proposal or construction.
Two common techniques are,
\begin{itemize}
\item Radiometric counting with high purity germanium (HPGe) gamma-ray spectrometers, and
\item Ion counting with inductively coupled plasma mass spectrometry (ICP-MS).
\end{itemize}

\subsection{High purity germanium assay (HPGe)}
\label{subsubsec:HPGe}

The concentrations of radioimpurities in a material sample can be measured by 
gamma ray spectrometry with HPGe detectors
through identification of the characteristic gamma-rays emitted from the decays of isotopes present in the material.

In a typical setup, 
a LN$_2$-cooled HPGe detector is housed in a chamber shielded from the ambient radiation by copper and lead. 
The sample to be assayed, usually 10 to 10$^3$ g in mass, is placed in the same chamber.
Gamma rays emitted by the radioimpurities in the sample deposit energy in the Ge crystal.

Background sources in the HPGe instrument include 
ambient radiation, 
cosmic ray induced energy deposits or activation,
and internal radioimpurities in the detector.
With sufficient shielding, ambient radiation is usually not a major source of background
and the effect of direct cosmic ray interactions can be reduced by using a muon veto or locating the detector underground. 
However, the effects of long-lived cosmogenic isotopes and internal radioimpurities cannot be easily removed.
This remaining background is measured through weeks-long background runs taken every several months.

Some of the gamma rays will deposit all of their energy in the Ge crystal, contributing to the so-called full peak.
The area of the full peak provides an estimate of the specific activity of the radioimpurity in question. 
More precisely,

\begin{equation}
\label{eq:geana}
X = \frac{1}{D} \cdot (s-b)
\end{equation}
where
$X$ is the specific activity of radioimpurity [Bq/kg];
$s$ and $b$ are the areas under the gamma ray spectra near the full peak 
in the sample run and the background run respectively [cps];
$D$ is the detection sensitivity [cps/(Bq/kg)] (aka efficiency), which equals $m\cdot f \cdot \epsilon(E)$ where
$m$ is the mass of the sample [kg]; 
$f$ is the branching fraction of the gamma ray; and 
$\epsilon(E)$ is the hit efficiency to be calibrated by detector simulation and/or 
radiation sources of known activity such as NIST traceable standards.
The error associated with $\epsilon(E)$ will not be considered in this study.

For this study whether to report a measured value or an upper limit is decided by 
the Feldman-Cousins (FC) approach \cite{FeldmanCousins} at a certain C.L., e.g. 90\%. 
We choose to use the FC method in this study because 
it is a readily reproducible prescriptive method available in the literature.
The reporting heuristic is:
if the lower limit evaluated by the FC approach equals zero, then an upper limit is reported 
in the form of ``$X < U_X$ at 90\% C.L.'', where

\begin{equation}
U_X = \frac{1}{D\cdot t} \mathrm{FC}^{90\%}_{\mathrm{UL}}\left( s \cdot t, \ b \cdot t \right)
\end{equation}
and $\mathrm{FC}^{90\%}_{\mathrm{UL}}$ 
is the upper limit evaluated by the FC approach at 90\% C.L., 
and $t$ is the HPGe run time.
Otherwise, a measurement value is reported in the form of ``$X \pm \sigma_X$'' where
$\sigma_X = \frac{1}{D}\sqrt{\frac{s + b}{t}}$,

\subsection{Inductively coupled plasma mass spectrometry (ICP-MS)}
\label{subsubsec:ICPMS}

While there are a variety of techniques for introducing samples to an ICP-MS, the vast majority of analyses introduce samples to the instrument in the form of a dissolved aqueous solution. The solution is nebulized into a fine aerosol in an argon carrier gas before introduction into the high temperature plasma (\textit{ca}. 6000-8000 K) where the dissolved components are atomized and ionized. The ions are directed through some ion optics in a high vacuum chamber before being mass resolved based on the mass-to-charge ratio of the ion in the mass analyzer and detected.

ICP-MS analyses use ``process blanks'' to account for the background contribution from the sample preparation steps prior to analysis (e.g., acid digestions, dilutions, purity from added reagents, etc.). The process blanks allow the sensitivity reach (detection limits) for the analysis to be determined.
In these ways ICP-MS has effectively the same underlying metrological and methodological requirements of HPGe counter, 
though they are implemented through differing means.

The sample and blank measurements are collected in units of counts per second.
The count rates can be converted to impurity concentration as follows,

\begin{equation}
\label{icpmstgt}
X = \frac{s}{D_s} - \frac{1}{n} \displaystyle\sum^n_i \frac{b_i}{D_{b_i}}
\end{equation}
where 
$s$ is the count rate for the analyte in the sample [$cps$];
$b_i$ is the count rate for the analyte in the $i$th blank [$cps$] out of a total of $n$ blanks;
$D_s$ and $D_{b_i}$ are the detection sensitivities [cps/ppt]. 

The detection sensitivities can be determined using a number of methods, but at PNNL, where we use isotope dilution methods, we determine the sensitivities using added non-naturally occurring tracers,

\begin{equation}
\label{icpmstrc1}
D_s = \frac{s'}{X'} 
\end{equation}
\begin{equation}
\label{icpmstrc2}
D_{b_i} = \frac{b_i'}{X'} 
\end{equation}
where 
the primed values are associated with the tracer. 
Here, for simplicity, we assume all the samples and blanks are spiked with the same amount of tracer.

Substituting equations \ref{icpmstrc1} and \ref{icpmstrc2}, equation \ref{icpmstgt} becomes,

\begin{equation}
X = X' \cdot \left(\frac{s}{s'} - \frac{1}{n} \displaystyle\sum^n_i \frac{b_i}{b'_i}\right)
\end{equation}

Despite the similarity to HPGe in terms of its counting statistics nature, 
the convention for reporting measurement results is somewhat different.
As an example, for ICP-MS measurements made at PNNL, the convention is:
if $s < 3 \sigma_b$, where $\sigma_b$ is the standard deviation of the ratios $\frac{b_i}{b'_i}$,
then $3 \sigma_b$ would be reported as the detection limit 
(approximately 95\% C.L. based on Student's t with 2 degrees of freedom);
otherwise, 
report $X \pm \sigma_X$ as the measurement result 
where 

\begin{equation}
\sigma_X = X'\cdot\sqrt{ \left(\frac{s}{s'}\right)^2 
                    \left[ \left( \frac{\sigma_s}{s} \right)^2 + \left(\frac{\sigma_{s'}}{s'}\right)^2 \right] + \sigma_b^2} 
\end{equation}
and $\sigma_s$ and $\sigma_{s'}$ are the errors on $s$ and $s'$ respectively.

\subsection{Summary of assay methodologies}

One basic difference between ICP-MS and HPGe counting is how 
variability is introduced by sample preparation.
Because the sample is typically introduced to the instrument in the form of a solution,
ICP-MS relies upon significantly more sample preparation steps than HPGe, 
typically requiring more added reagents and sample digestion steps.
However, ICP-MS measurements are buttressed with process blank preparations and measurements 
along with the most accurate and precise quantitation method for ICP-MS by employing isotope dilution
\cite{IcpmsMethod,IcpmsMethod2}.
HPGe counting minimizes sample preparation, 
but at the cost of assumptions made regarding the background stability of the counting apparatus and 
the appropriateness of calibration standards to be representative of the sample being measured.
The concern for variability in either case, ICP-MS or HPGe, is 
whether the statistical distribution for repeated measurements obeys Poisson statistics.
This can be tested, via good laboratory practice, in both cases. 
For the purposes of this study, 
we presume such good laboratory practice is achieved 
as the goal of this study is to understand 
how the different treatments of assay results
impact sensitivity projections 
rather than to evaluate the assay methodologies.

Another point to note is that the two methods typically have very different detection sensitivities.
ICP-MS can reach sub-ppt detection ($\mu$Bq/kg) levels 
with tens of milligrams of sample material.
In comparison, HPGe is typically 2-3 orders of magnitude less sensitive
than ICP-MS, even though hundreds of grams of sample material is typically used. 
However, the magnitude of detection sensitivity has no direct effect on the statistical distribution of repeated measurements.


Time variation in detection sensitivity $D$ is known to exist in 
both HPGe \cite{Knoll} and ICP-MS \cite{IcpmsPoisson}.
For HPGe, 
detection sensitivity $D$ 
is periodically calibrated with button sources;
while for ICP-MS, \textit{in situ} tracers are used to estimate $D$ for every run.
These methods establish an \textit{average} $D$ for a run. 
Therefore the treatment of time variation in both methods are mathematically identical.

Another difference is that HPGe and ICP-MS count different species.
HPGe counts the gamma rays from the radioactive decays of the impurity and/or its progenies,
while ICP-MS counts the ions of the parent isotope of the decay chain. 
The results from the two techniques can be compared
when the impurity in the sample is in secular equilibrium.
In other words, under this assumption, 
the measurement results of HPGe and ICP-MS can be seen as the same quantity merely in different units (Bq/kg vs ppt), 
as can their detection sensitivities $D$ (cps/(Bq/kg) vs cps/ppt).

To sum up, despite the apparent differences discussed above,
with good laboratory practice and the assumption of secular equilibrium, 
whether a measurement is made by HPGe counting or ICP-MS has no effect on its statistical properties
relevant to this study.

\section{Model of assay, experiment, and sensitivity}
\label{sec:Model}

\subsection{Hypothetical experiment}
\label{subsec:Experiment}
Dark matter and neutrinoless double beta decay experiments ($0\nu\beta\beta$) are 
the primary large, low-background experiments 
facing the issue of treating \textit{upper limits} in their material assay results 
when modeling their experimental sensitivity. 
We choose to focus on a simplified model of a hypothetical $0\nu\beta\beta$ decay experiment 
to further explore the impact of treating material assay \textit{upper limits}. 
This choice benefits from the signature of the $0\nu\beta\beta$ decay: 
A search for an energy peak (typically Gaussian) at the known $Q$-value, $Q_{\beta\beta}$, of the decay. 
For this reason our hypothetical experiment can largely be treated as a counting experiment, 
closely aligned with the earlier discussions within this paper regarding counting statistics. 
This is in contrast to employing a hypothetical model of a dark matter experiment, 
where we would need to bring in details of the dark matter hypothesis that are unrelated to the focus of this paper.

\paragraph{Detector}
Our hypothetical $0\nu\beta\beta$ detector consists of $N$ parts with 
masses $M_i$, 
impurity concentrations $X_i$ and 
hit efficiencies $\varepsilon_i$. 
The fiducial volume contains one metric ton of radiopure $^{136}$Xe.\footnotemark
\footnotetext{
A specific isotope, here $^{136}$Xe, is chosen just for concreteness.
 This study does not depend on the isotope choice.} 
The impurity concentrations $X_i$ are assumed to be uncorrelated.
The livetime of this hypothetical experiment is assumed to be 10 years.

As seen in Equation \ref{eq:SimpleRate}, since $M_i$ and $\varepsilon_i$ always appear as a product, 
for the purpose of sensitivity projection,
we can assume $M_i=1$ for all $i$ without loss of generality, 
as the variation in mass can be absorbed into $\varepsilon_i$.
Therefore, the detector can be completely specified by two one-dimensional arrays, $\{X_i\}$ and $\{\varepsilon_i\}$ 
(with $N$ implicitly specified by the size of the arrays). 

\paragraph{Assay}

Given the discussion of Section \ref{subsec:AssayMethods} and a desire for simplicity
we assume that all materials are assayed with an HPGe detector. 
We further assume that, in the materials, there is only one relevant radioimpurity
whose concentration is estimated by integrating the HPGe measured counts
in a narrow region of interest around 
the spectral peak at the characteristic energy of the one gamma-ray
that it emits. 
For ease and no loss of generality towards our purpose,
the detection efficiency for the radioisotope is assumed to be 100\%.
The HPGe detector is assumed to have a background rate of 10 cpd in the region of interest.
All material samples used for assay are one kg, and are counted for 14 days. 
The background runs are also 14 days long and each background run is paired with only one sample run.

An assay measurement by the HPGe detector is simulated by drawing two Poisson random variables, 
representing the sample counts and the background counts.
Results are analyzed and reported following the heuristic described above in Section \ref{subsubsec:HPGe}.
The central values (negative or not) and their uncertainties are always reported, 
so that they can potentially be used in defining a prior.

Under these assumptions, 
the detection sensitivity $D$ is $10^{-6}$ cps/($\mu$Bq/kg), 
and the background rate $b$ is 10 cpd (or $1.16\times10^{-4}$ cps).
It can be shown that a sample with impurity concentration of $X=23.8 ~\mu$Bq/kg 
would be barely detectable by this detector. (See \ref{sec:hpgesens} for derivation.)

\subsection{Sensitivity method}
\label{subsec:Sensitivity}

\begin{figure}[htbp]
\centering
\includegraphics[width=1.1\textwidth]{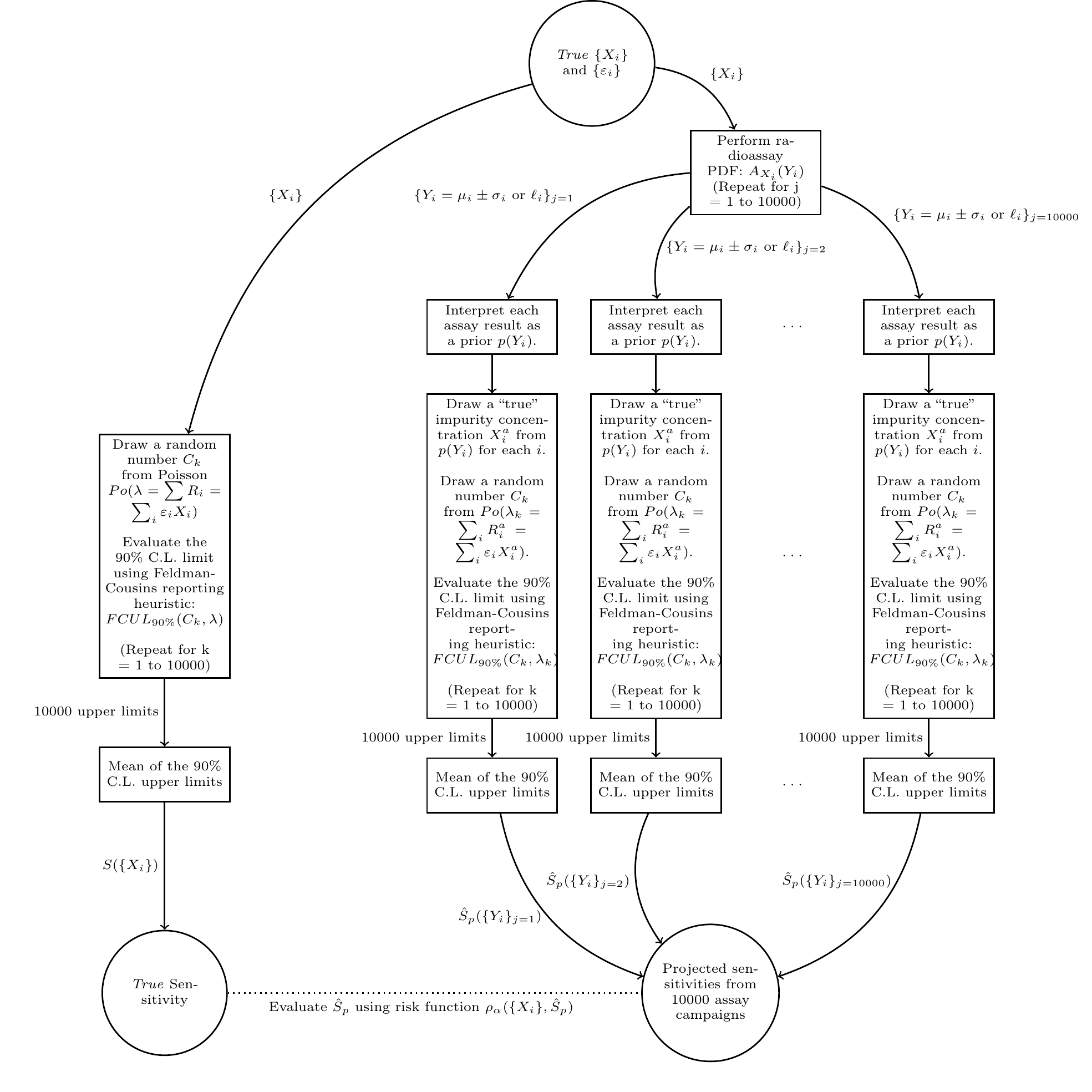}
\caption{
Diagram of the implementation of the sensitivity calculation.
Here $\{X_i\}$ is the set of the \emph{true} impurity levels for all the detector parts (of index $i$.)
Likewise $\{Y_i\}$ is a set of the impurity levels for all detector parts 
\emph{derived} from a model of the radioassay measurement process. 
The index, $j$, represents the imagining performing all radiopurity assays multiple times, 
with each case representing one possible way the radioassay program might report 
derived impurity levels $\{Y_i\}$ based on the \textit{true} impurity levels $\{X_i\}$.}
\label{fig:senscalc}
\end{figure}

In a nutshell, the projected sensitivity is calculated by a Monte Carlo method \footnotemark,
in which a large number of toy experiments are simulated under the null hypothesis.
Each toy represents a possible realization of the experiment, where an upper limit on the signal can be set.
The mean of these upper limits is the rate sensitivity which can be converted to a half-life sensitivity. 
(See \ref{sec:avgsens} for discussion.)
The calculation procedures are detailed below, also illustrated in Figure \ref{fig:senscalc}.
\footnotetext{
The sensitivity projection code is available at: \url{https://github.com/pnnl/sensitivity}
}

\subsubsection{Sensitivity based on perfect knowledge}
If the actual \textit{true} impurity concentrations of the detector parts are known (as we can in our idealized model),
the expected background count ($\lambda$) can be calculated precisely as $\lambda = \sum_i R_i = \sum_i \varepsilon_i X_i$,
and the background counts realized in the toys ($C_k$) are drawn 
from the Poisson distribution, $Po(\lambda)$.
The following procedure is repeated 10000 times (numbering, k) 
to produce an ensemble of possible outcomes for the toys:

\begin{enumerate}
\item Draw a random number from $Po(\lambda)$, and assign it to $C_k$.
\item Calculate the 90\% C.L. upper limit for an experiment with expected background count $\lambda$ 
      and measured count  $C_k$ using the FC reporting heuristic.
\item Append the upper limit to an array used to collect the ensemble of outcomes for the toys.
\end{enumerate}

In this case, as we have perfect knowledge of the impurity levels in this hypothetical study,
the \textit{true} half-life sensitivity ($S$) is calculated as

\begin{equation}
S = N_{Xe} \ln(2)/S_R 
\end{equation}
where $N_{Xe}$ is the number of $^{136}$Xe nuclei in the fiducial volume, and 
$S_R$ is the mean of the upper limits saved to the array.
The above described procedure corresponds to the leftmost flow from top to bottom in the diagram in Figure \ref{fig:senscalc}.
Notice that $S$ is a fixed number despite the use of Monte Carlo method in its calculation.

\subsubsection{Sensitivity based on radioassay measurements}

\begin{figure}[!htbp]
\centering
  \begin{subfigure}[t]{.48\textwidth}  
    \includegraphics[width=\textwidth]{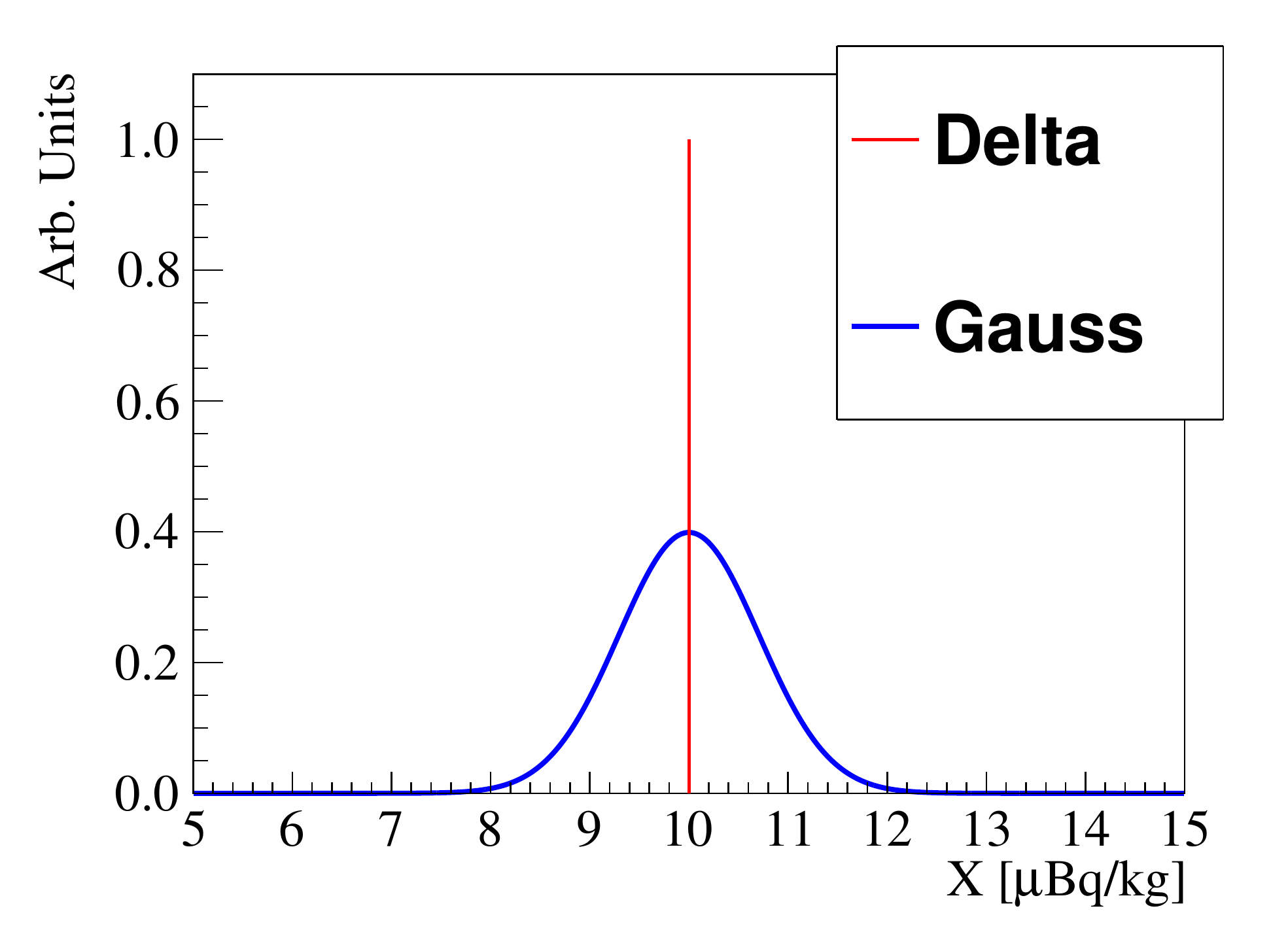}
    \caption{Central value: $X = (10 \pm 1)$ $\mu$Bq/kg}
    \label{fig:policies-cv}
  \end{subfigure}
  \begin{subfigure}[t]{.48\textwidth}
    \includegraphics[width=\textwidth]{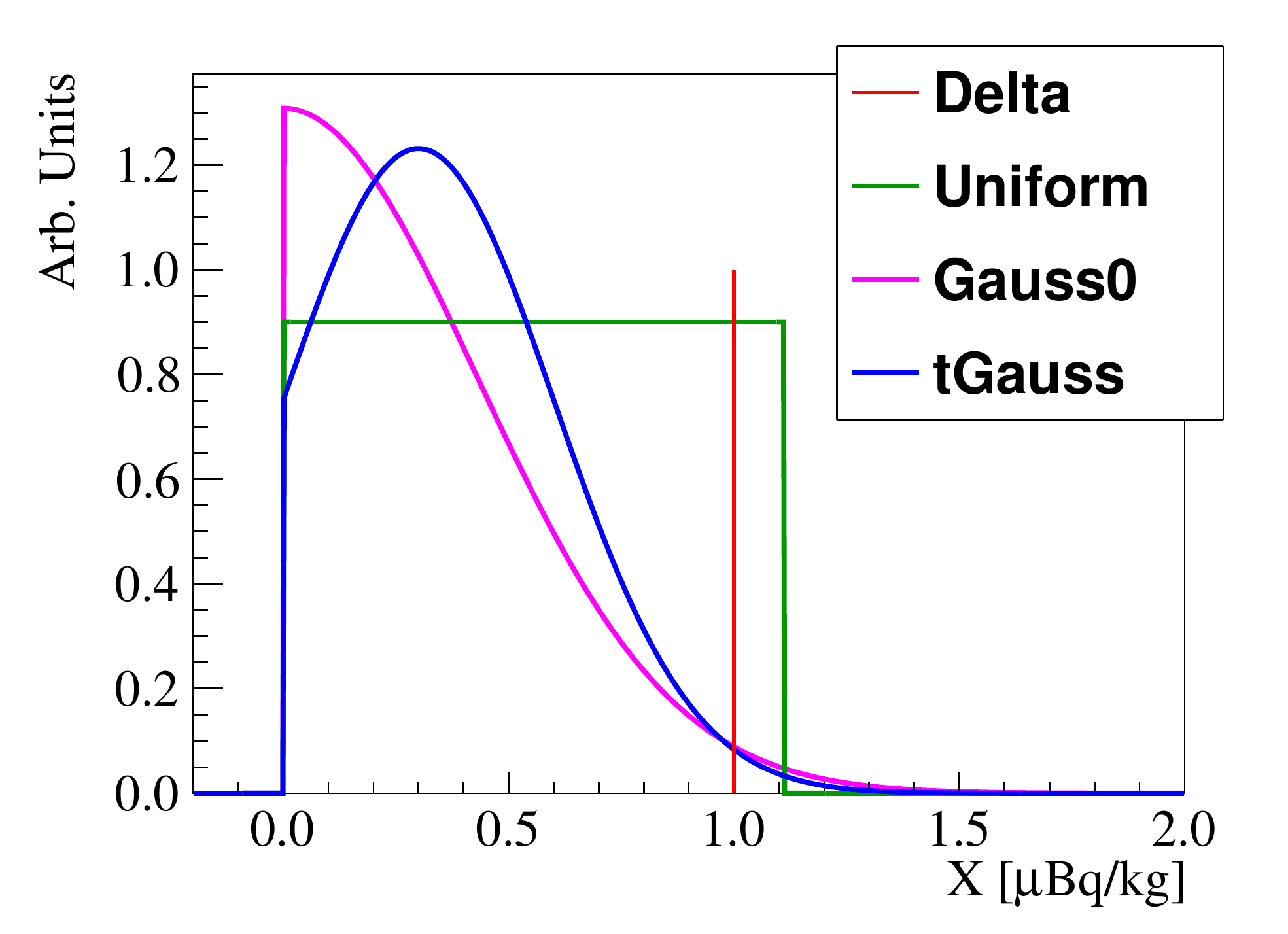}
    \caption{Limit: $X < 1$ $\mu$Bq/kg (or $X = (0.30 \pm 0.43)$ $\mu$Bq/kg)}
    \label{fig:policies-limit}
  \end{subfigure}
\caption{\label{fig:policies}
Illustrations of prior choices considered. (Colors online)
}
\end{figure}

Impurity concentrations measured in material assays are often reported in two different forms, 
either as a central value with an error ($\mu_i \pm \sigma_i$) or as an upper limit ($\ell_i$).
The following details several common ways to use them in a sensitivity calculation, 
as illustrated in Figure \ref{fig:policies}.

\paragraph{Handling of central values}

We consider two possible choices. 
Both of which can be expressed in terms of drawing a ``true'' \footnotemark impurity concentration from a prior.
\footnotetext{\textit{True} (in italics) represents the fixed but unknown value of a parameter in frequentist interpretation, 
whereas ``true'' (in quotes) represents a possible value of a parameter drawn from a Bayesian prior.}
\begin{enumerate}
\item \textbf{Dirac delta prior: $\delta(x-\mu_i)$}
When the central value is large compared to the error, it is often used directly
for the impurity concentrations in the sensitivity calculation, as if we have perfect knowledge.
\item \textbf{Gaussian prior: $G(x;\mu_i,\sigma_i)$}
However, when the central value is comparable to the measurement uncertainty
(``just a few sigmas away from zero'' for example), 
the knowledge of the precision of the impurity concentration may be captured with a prior describing the assay result.
The ``true'' impurity concentration ($X_i^a$) is then drawn from such a prior.
Arguably the most intuitive choice is 
a Gaussian with $\mu$ equals the central value and $\sigma$ equals the measurement uncertainty ($G(\mu_i,\sigma_i)$).
Note that this prior choice is also motivated by Bayesian arguments (as described in \ref{sec:bayes}.)

\end{enumerate}
 
\paragraph{Handling of upper limits}

When an upper limit is reported, we do not know
the most likely value of the impurity concentration.
Below are some possible ways to cope with this situation
using assumed priors to once again capture the (now more limited) knowledge gained from the radioassay measurement.

\begin{enumerate}
\item \textbf{Dirac delta prior $\delta(x-\ell_i)$}.
The upper limit is used for the ``true'' impurity concentration. 
As one might expect, the calculated sensitivity could be worse (or more ``conservative'') 
than the actual \textit{true} sensitivity, 
as we will show in Section \ref{sec:Discussion}.
Also, this method does not respect the confidence level of the reported upper limit, if it is associated with one.
\item \textbf{Uniform prior $U_{(0,\ell_i/0.9)}(x)$}.
Another approach is to assume no further knowledge than is given by the upper limit itself. 
In other words, the ``true'' impurity concentration is considered equally likely to lie
anywhere between 0 and the reported upper limit.
As a technical detail, 
the uniform prior is defined to extend beyond the reported upper limit to preserve the confidence level of 90\%.
\item \textbf{Gaussian prior}
A Gaussian prior seems a natural choice
to represent the ``true'' impurity concentration,
but the challenge here is to set both parameters $\mu$ and $\sigma$ 
when only one value (the upper limit $\ell_i$) is available.
We consider two possible strategies:
\begin{enumerate}
\item \textbf{Half-Gaussian prior centered at zero (Gauss0)}
$G(x; 0, \ell_i/1.64) $
The reporting of an upper limit implies that the actual \textit{true} impurity concentration could be zero. 
One could posit that the most likely value is indeed zero and 
the non-observation of impurities is evidence for such a state of affairs.
To express this belief,
a prior would thus be formed with a half-Gaussian with $\mu$ equals 0 and $\sigma$ adjusted 
so that the tail area 
conforms with the confidence level of the reported limit
(i.e. $\int_{\ell_i}^{\infty} G(x;0,\sigma)~dx = 0.1$, if the C.L. is assumed to be 90\%).
\item \textbf{Truncated Gaussian (tGauss)}
$G(x; \mu_i, \sigma_i) $
If we have access to the measurement counts that beget the limit, there is another option
motivated by Bayesian arguments (as elaborated in \ref{sec:bayes}.)
If we set $\mu$ to be $\frac{1}{D}\cdot (s-b)$ (as in Equation \ref{eq:geana}) which may be negative, 
and set $\sigma$ to the usual measurement uncertainty, 
we can then form a Gaussian. This Gaussian will be truncated at zero to remove unphysical impurity concentrations.
However, this is not always feasible as this requires $\mu_i$ and $\sigma_i$ while usually only $\ell_i$ is reported.
This last point will be discussed further in Section \ref{sec:Discussion}.
\end{enumerate}
\end{enumerate}

\section{Projected sensitivity under different scenarios}
\label{sec:Scenarios}

\subsection{Detector with identical parts}

First consider a detector consisting of $N$ identical parts, 
each part having a hit efficiency of 
$\varepsilon = 3.171\times10^{-4}/N$ 
so that the \textit{true} total background rate is 1 count per year when $X = 100$ $\mu$Bq/kg.
Recall that $X = 100$ $\mu$Bq/kg is well above the detection limit of our HPGe assay detector ($\sim$$23.8~\mu$Bq/kg.)

\begin{figure}[htbp]
  \centering
  \begin{subfigure}[t]{.48\textwidth}
    \includegraphics[width=\textwidth]{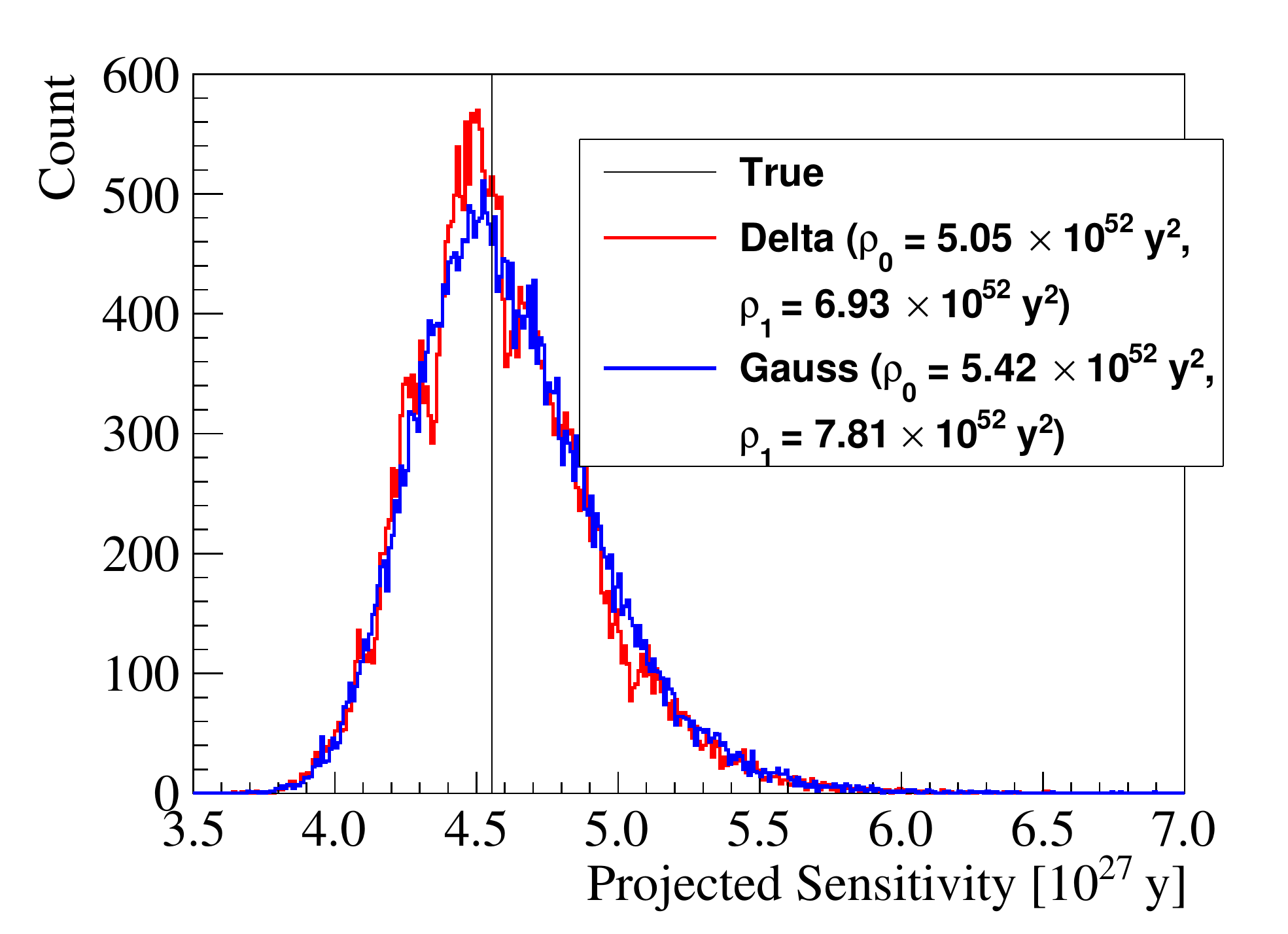}
    \caption{N = 1}
    \label{fig:deltavsgauss-N1}
  \end{subfigure}
  \begin{subfigure}[t]{.48\textwidth}
    \includegraphics[width=\textwidth]{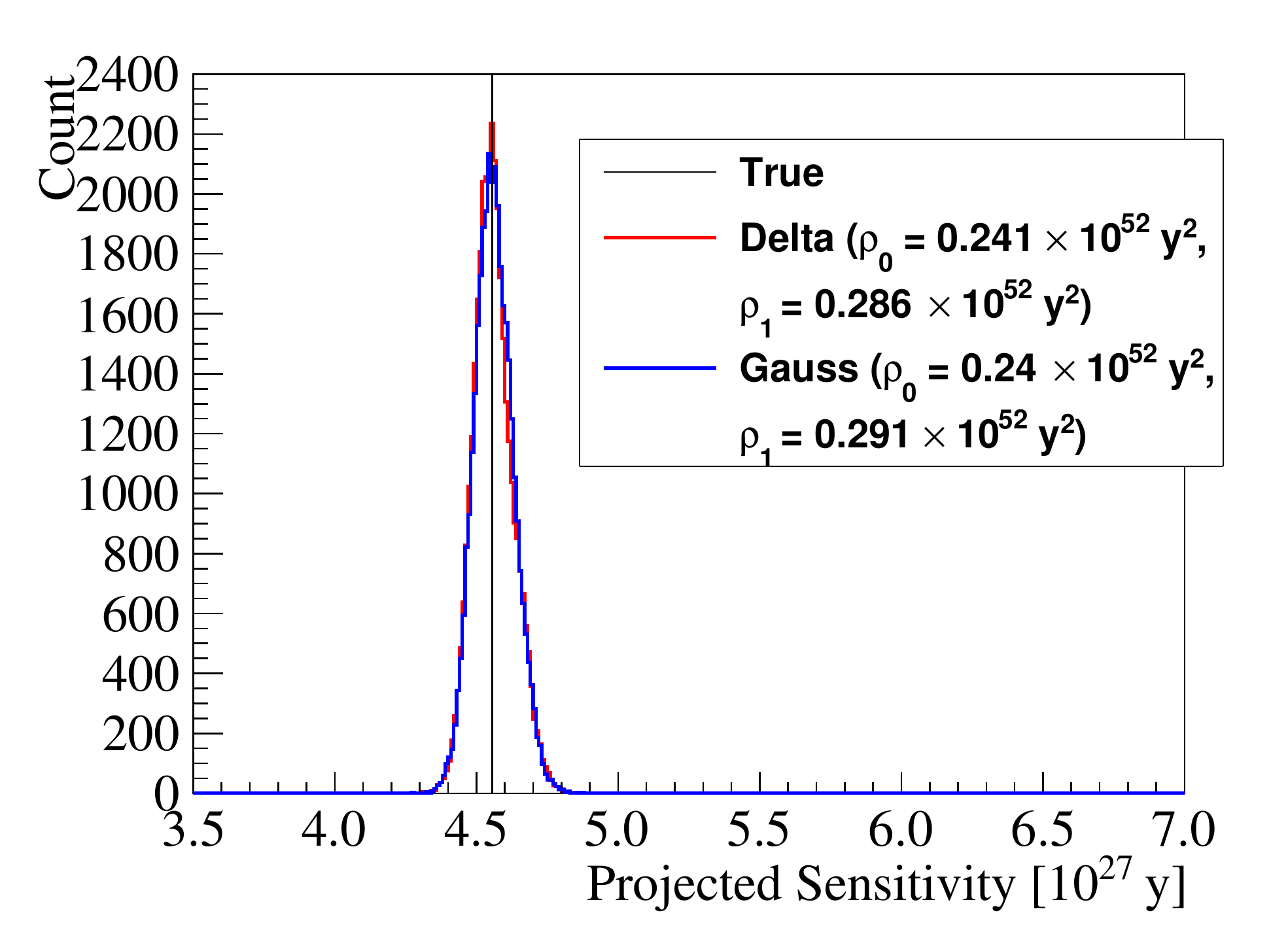}
    \caption{N = 20}
    \label{fig:deltavsgauss-N20}
  \end{subfigure}
  \caption{Comparison between the Dirac delta prior and the Gaussian prior for assay measurements 
    reporting central values and errors, assuming detector with $N$ identical parts, 
    each part having impurity concentration X = 100 $\mu$Bq/kg, 
    and $\varepsilon = 3.171 \times 10^{-4}/N$. 
    The reduction in spread for large $N$ can be attributed to the independence of the assays performed for the $N$ parts.
    Each histogram contains 35000 realizations of assay campaigns.
    The risks $\rho_0$ and $\rho_1$ are defined in Section \ref{sec:Discussion}. (Colors online)
    }
  \label{fig:deltavsgauss}
\end{figure}

When $X = 100$ $\mu$Bq/kg,
only the prior choice for central values is relevant as
the probability of reporting an upper limit is vanishingly small.
Figure \ref{fig:deltavsgauss} compares the projected sensitivities 
for using the Dirac delta prior and the Gaussian prior 
in such a case,
for $N = 1$ and $N = 20$.
As seen in the figures,
the difference between the two priors is relatively small (as compared to the results that follow).
For ease of comparison, in all following calculations,
the Dirac delta prior is always used if a central value is reported by the HPGe assay instrument.

\begin{figure}[htbp]
  \begin{subfigure}[t]{.5\textwidth}
    \includegraphics[width=\textwidth]{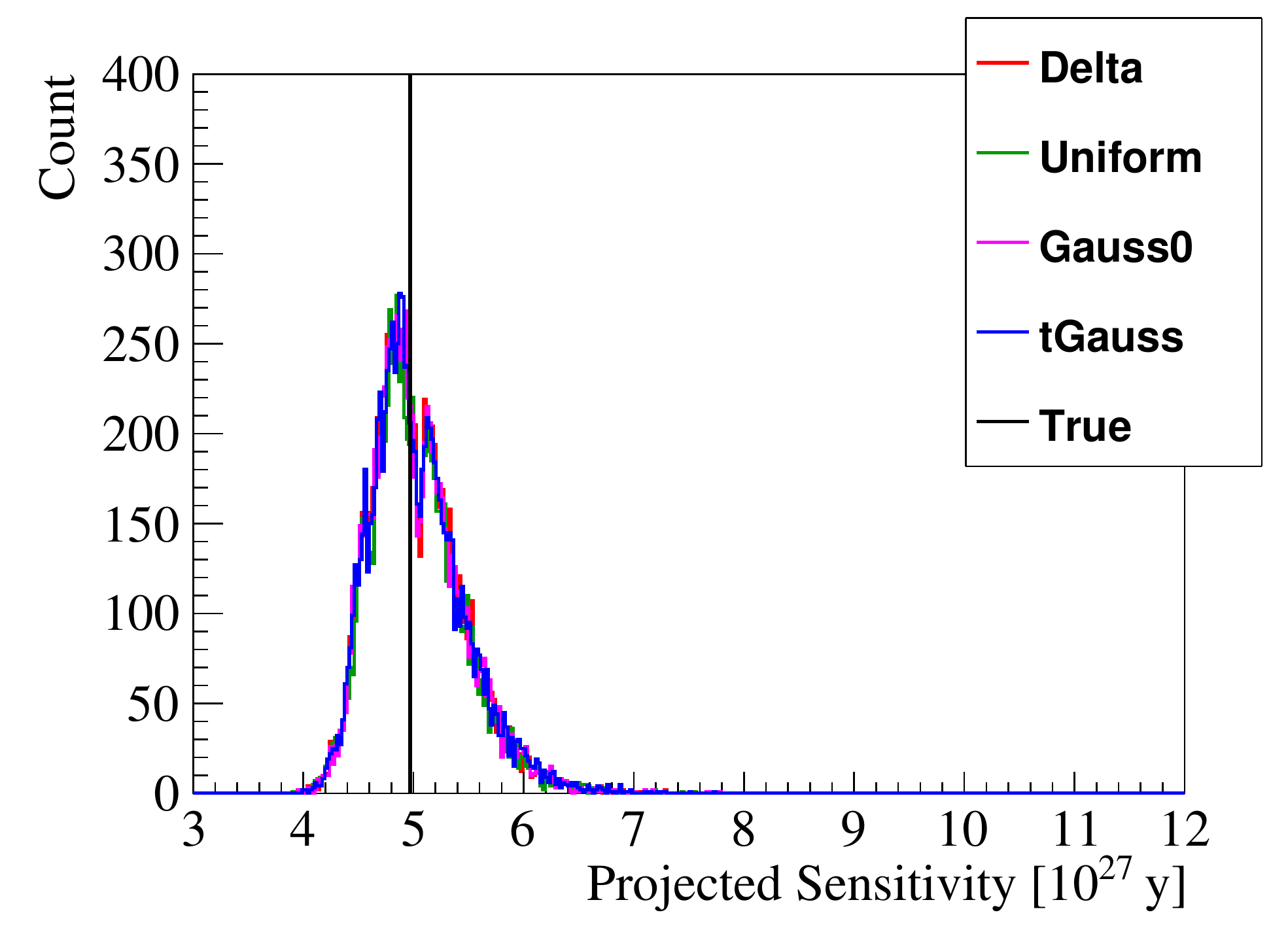} 
    \caption{$X = 80 ~\mu$Bq/kg}
    \label{fig:DeltaPriors-N1-R8}
  \end{subfigure}
  \begin{subfigure}[t]{.5\textwidth}
    \includegraphics[width=\textwidth]{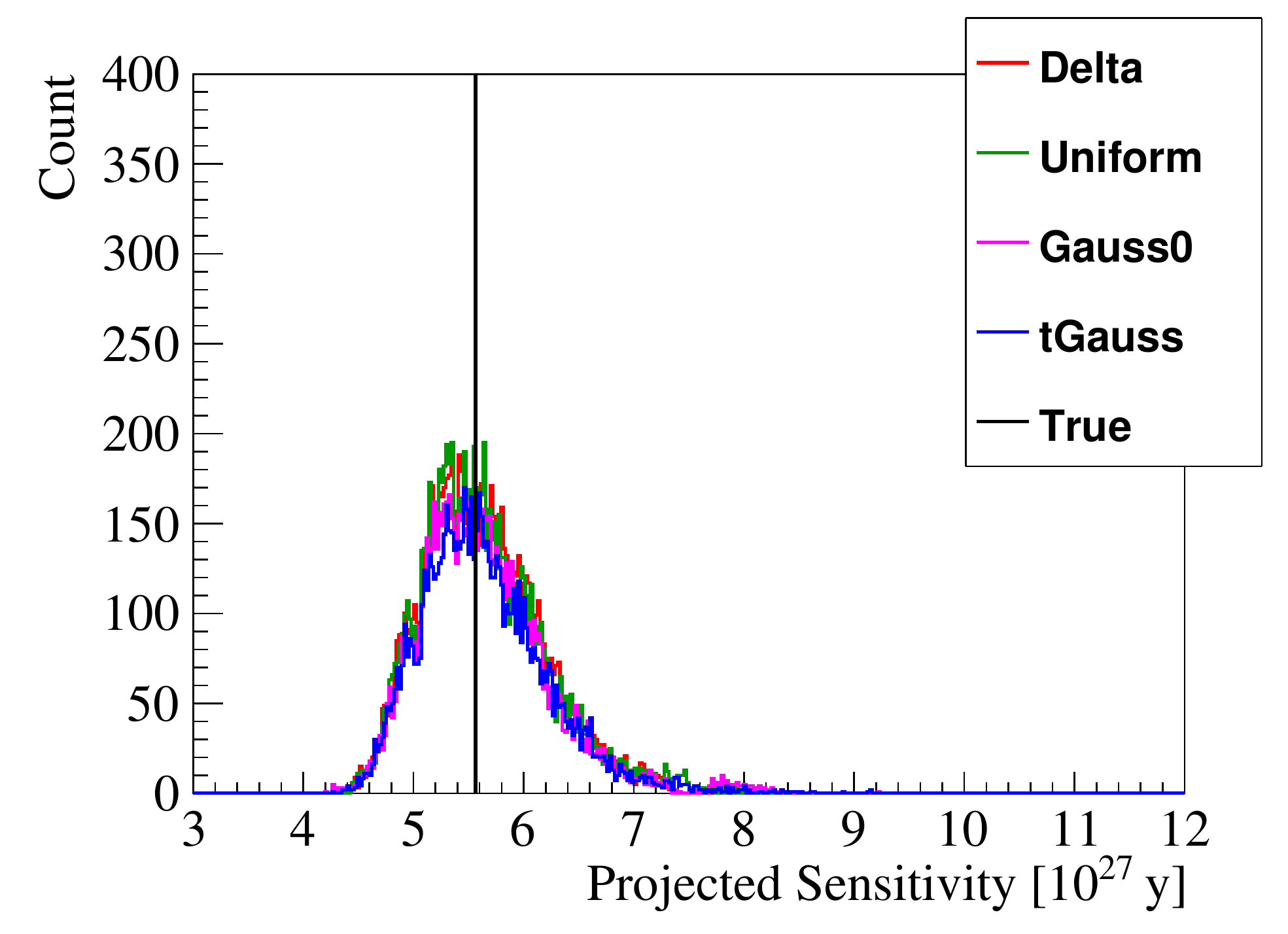} 
    \caption{$X = 60 ~\mu$Bq/kg}
    \label{fig:DeltaPriors-N1-R6}
  \end{subfigure}
  \begin{subfigure}[t]{.5\textwidth}
    \includegraphics[width=\textwidth]{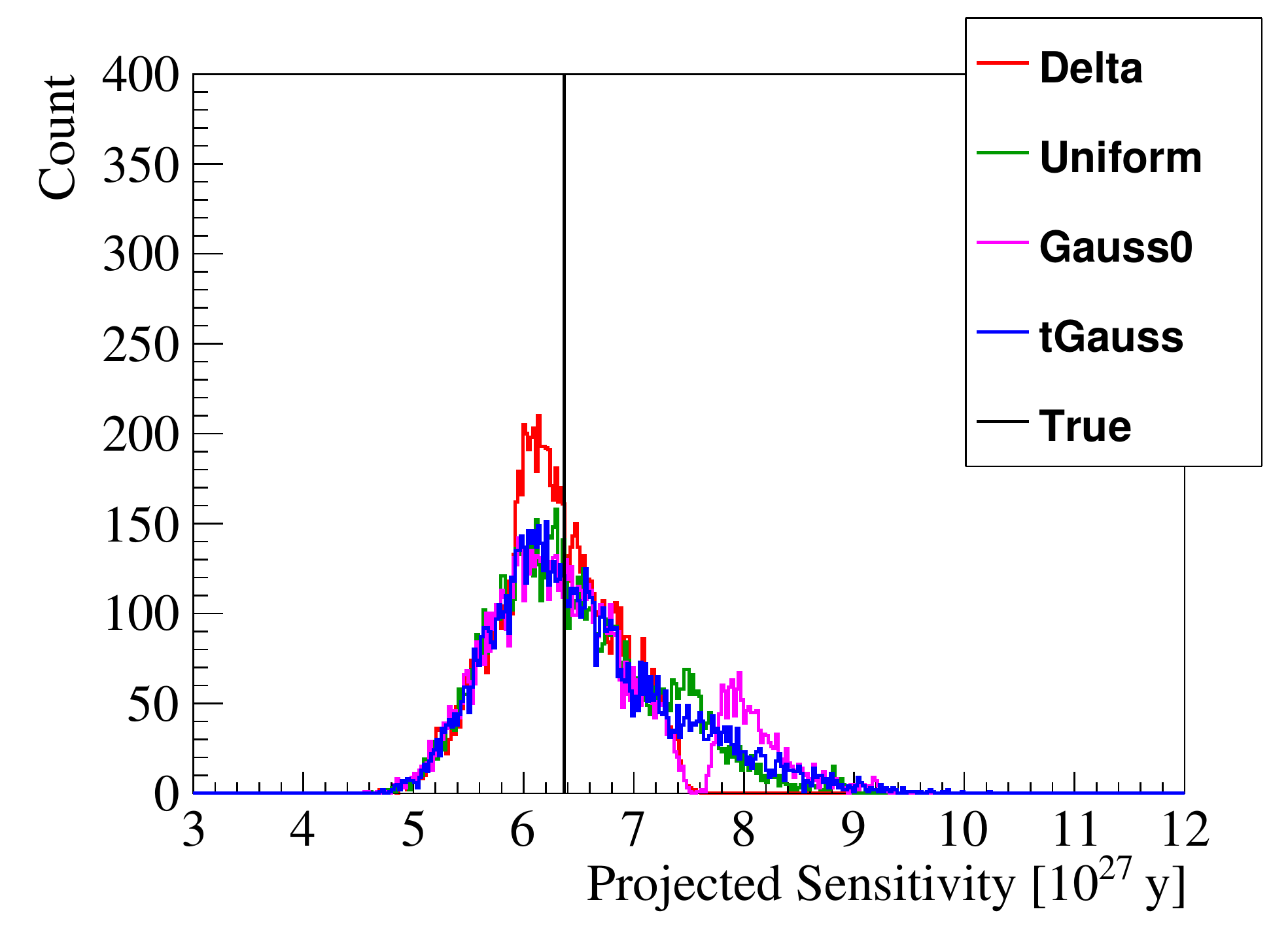} 
    \caption{$X = 40 ~\mu$Bq/kg}
    \label{fig:DeltaPriors-N1-R4}
  \end{subfigure}
  \begin{subfigure}[t]{.5\textwidth}
    \includegraphics[width=\textwidth]{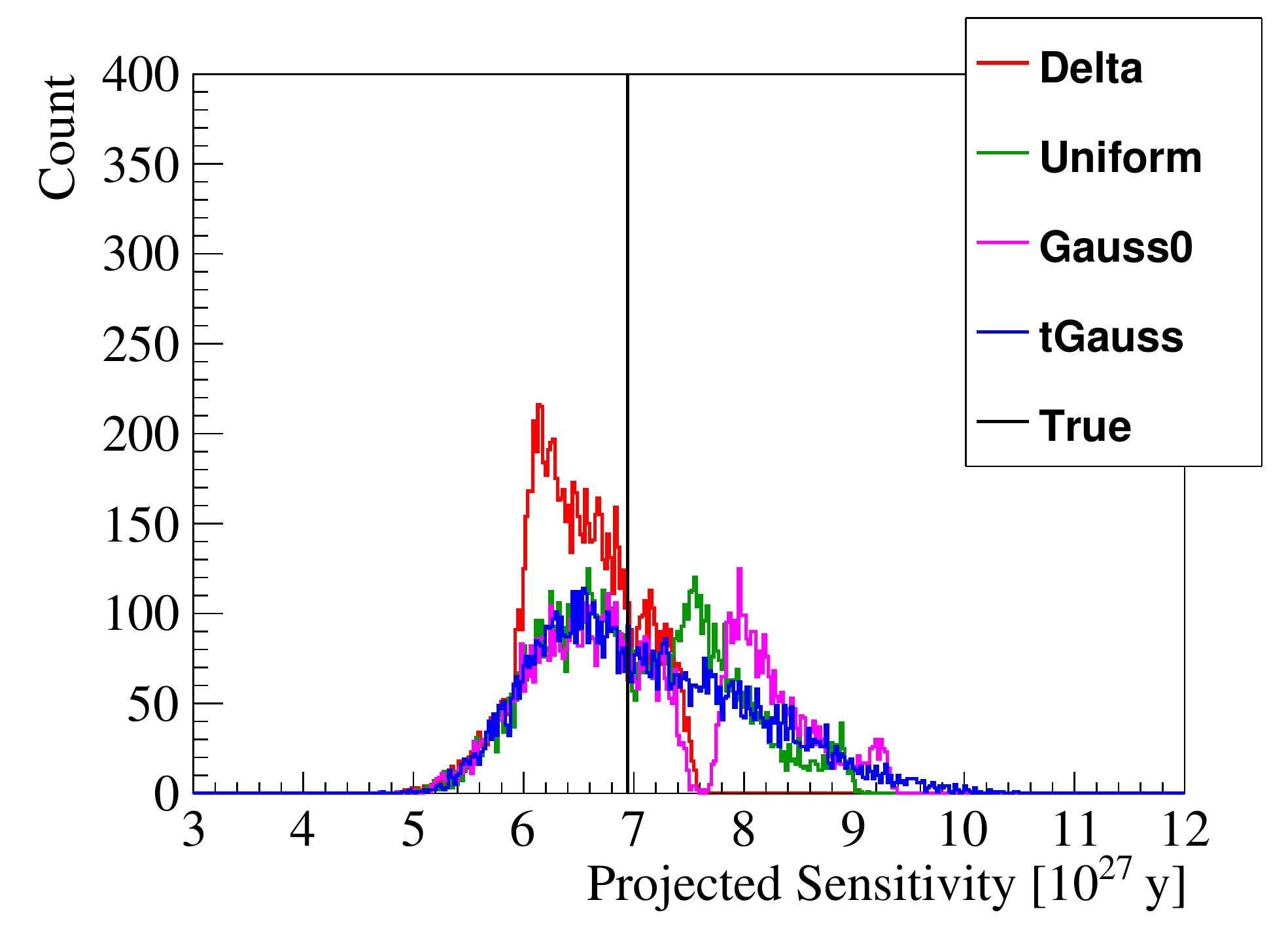} 
    \caption{$X = 30 ~\mu$Bq/kg}
    \label{fig:DeltaPriors-N1-R3}
  \end{subfigure}
  \begin{subfigure}[t]{.5\textwidth}
    \includegraphics[width=\textwidth]{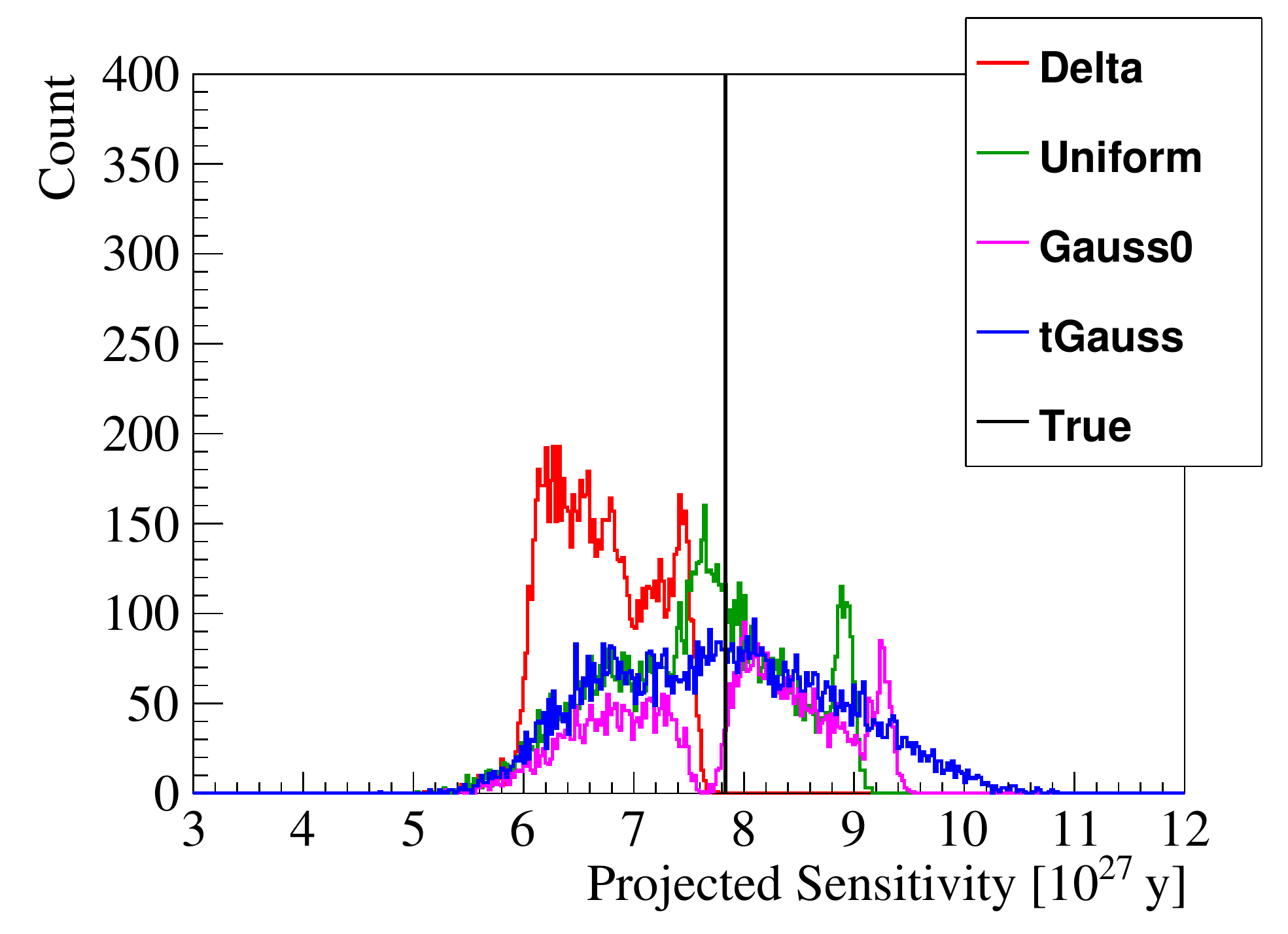} 
    \caption{$X = 20 ~\mu$Bq/kg}
    \label{fig:DeltaPriors-N1-R2}
  \end{subfigure}
  \begin{subfigure}[t]{.5\textwidth}
    \includegraphics[width=\textwidth]{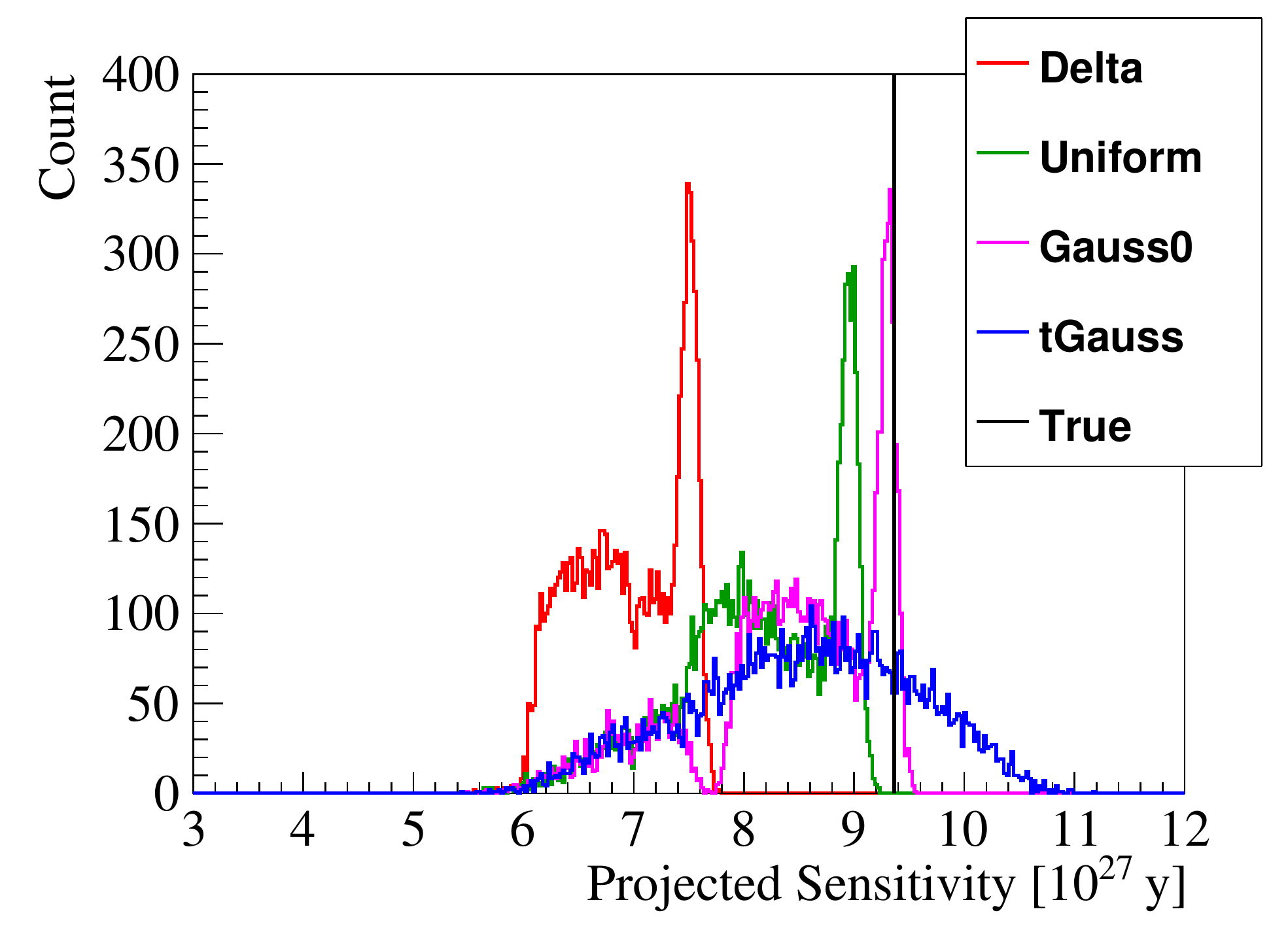} 
    \caption{$X = 10 ~\mu$Bq/kg}
    \label{fig:DeltaPriors-N1-R1}
  \end{subfigure}
  \caption{Projected sensitivities for different priors and different impurity concentrations $X$, 
for $N = 1$ and $\varepsilon = 3.171\times10^{-4}$. 
Notice that panel (a) having $X$ = 80 $\mu$Bq/kg is very similar to Figure \ref{fig:deltavsgauss-N1} 
where $X$ = 100 $\mu$Bq/kg. 
As the impurity concentration decreases toward $X$ = 10 $\mu$Bq/kg (panel (f)), 
three of the four choices of prior show similar but highly-asymmetric distribution shapes. (Colors online) }
  \label{fig:DeltaPriors-N1}
\end{figure}

\begin{figure}[htbp]
  \begin{subfigure}[t]{.5\textwidth}
    \includegraphics[width=\textwidth]{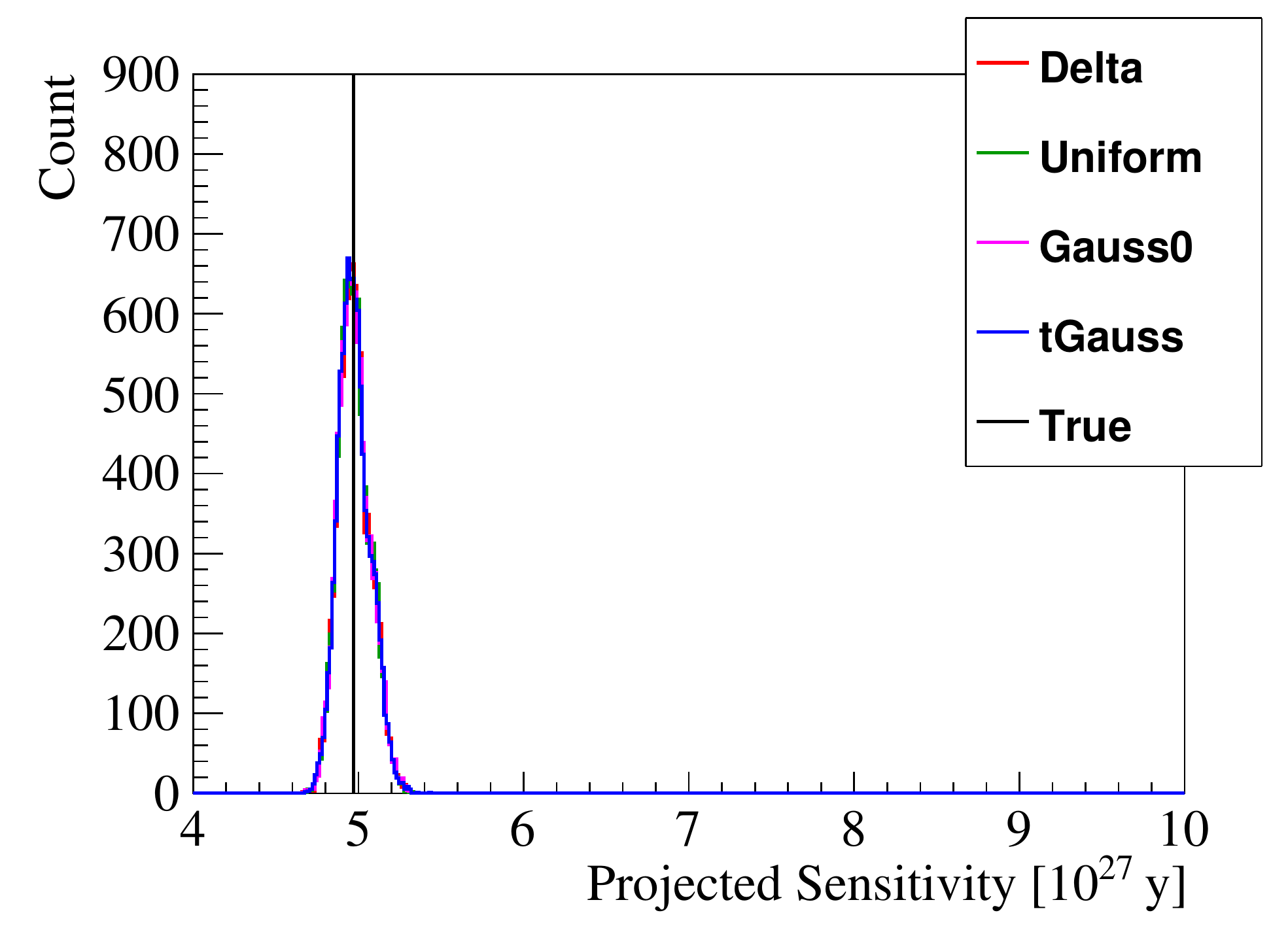} 
    \caption{$X = 80 ~\mu$Bq/kg}
    \label{fig:DeltaPriors-N20-R8}
  \end{subfigure}
  \begin{subfigure}[t]{.5\textwidth}
    \includegraphics[width=\textwidth]{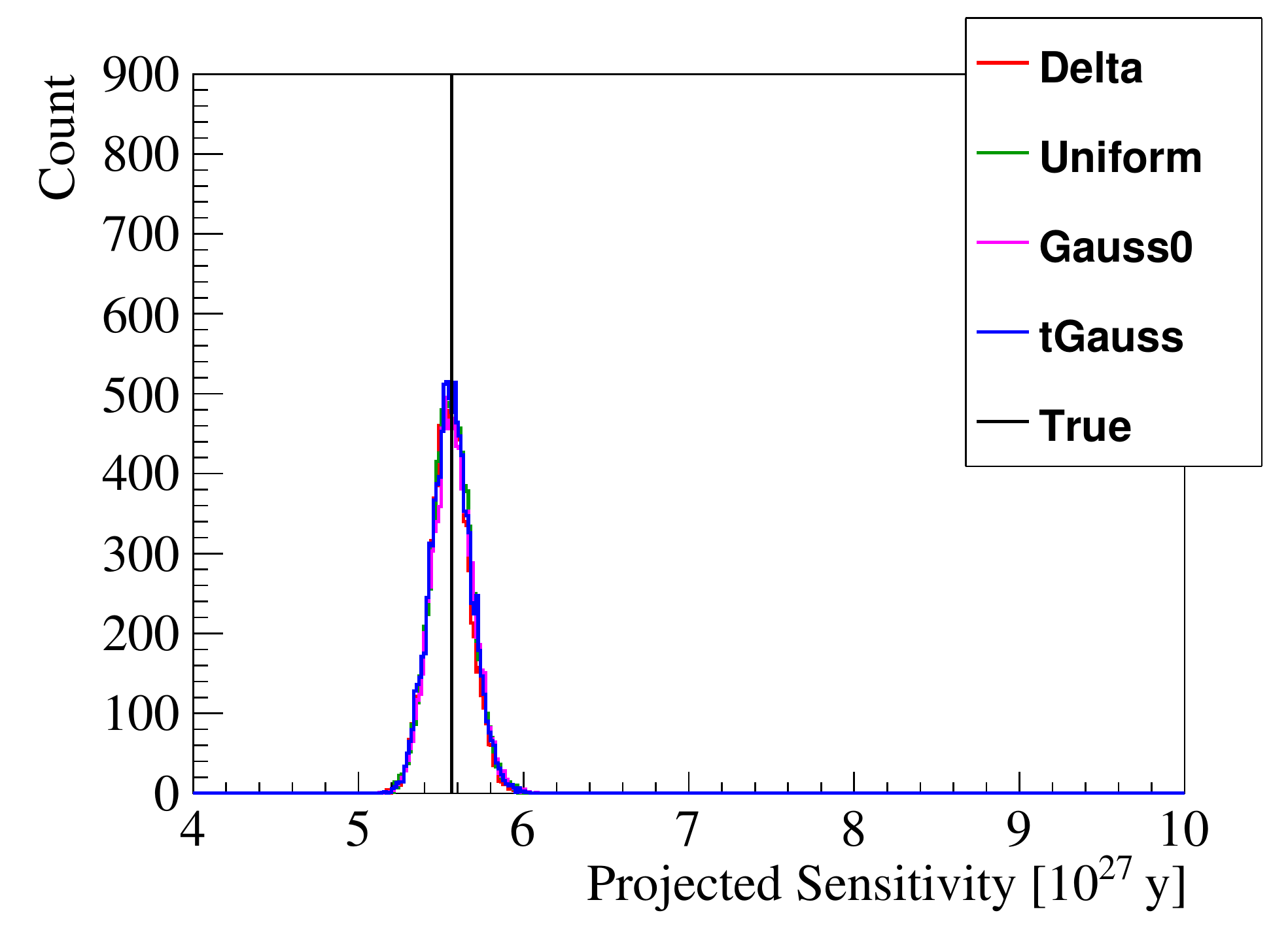} 
    \caption{$X = 60 ~\mu$Bq/kg}
    \label{fig:DeltaPriors-N20-R6}
  \end{subfigure}
  \begin{subfigure}[t]{.5\textwidth}
    \includegraphics[width=\textwidth]{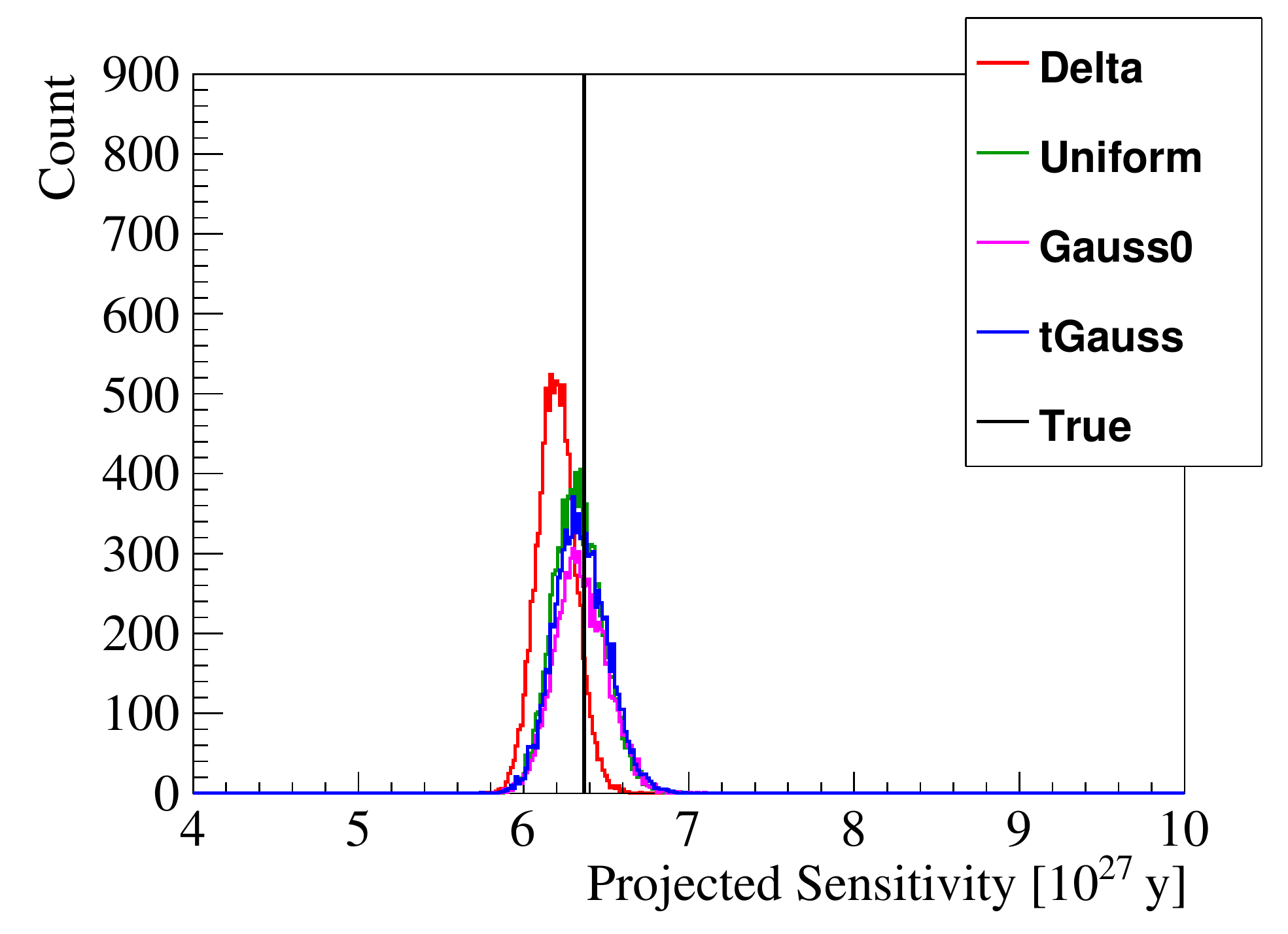} 
    \caption{$X = 40 ~\mu$Bq/kg}
    \label{fig:DeltaPriors-N20-R4}
  \end{subfigure}
  \begin{subfigure}[t]{.5\textwidth}
    \includegraphics[width=\textwidth]{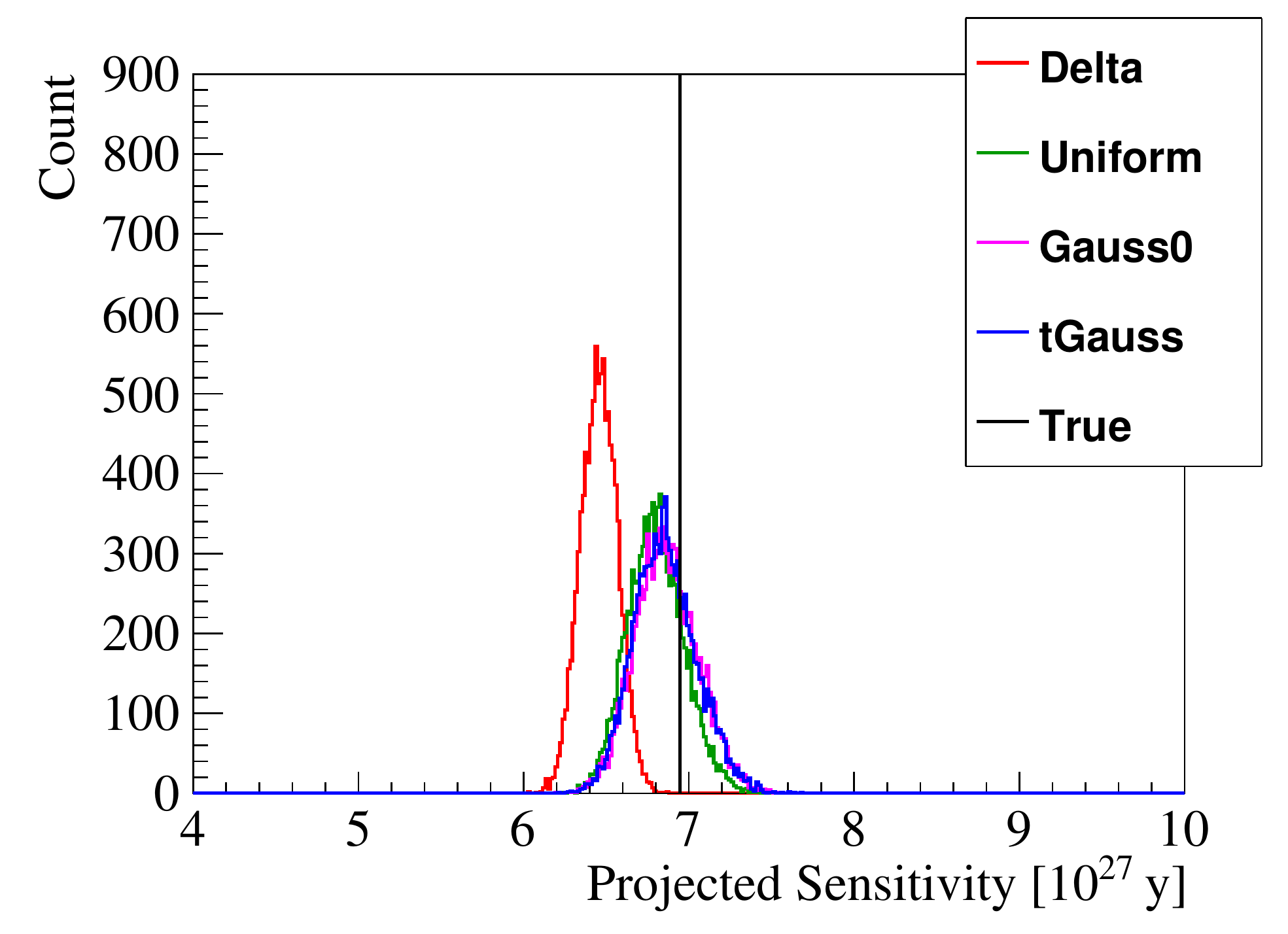} 
    \caption{$X = 30 ~\mu$Bq/kg}
    \label{fig:DeltaPriors-N20-R3}
  \end{subfigure}
  \begin{subfigure}[t]{.5\textwidth}
    \includegraphics[width=\textwidth]{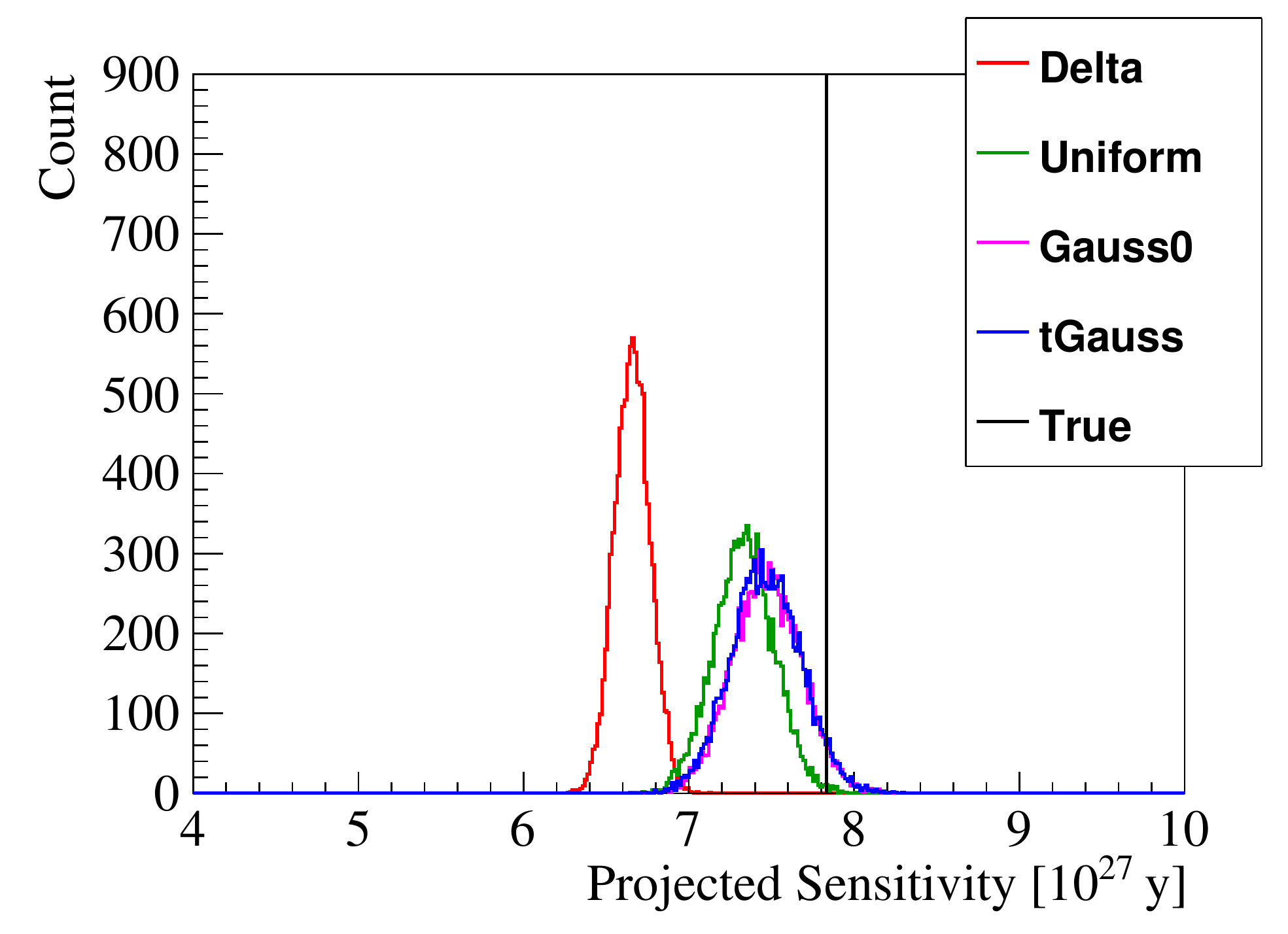} 
    \caption{$X = 20 ~\mu$Bq/kg}
    \label{fig:DeltaPriors-N20-R2}
  \end{subfigure}
  \begin{subfigure}[t]{.5\textwidth}
    \includegraphics[width=\textwidth]{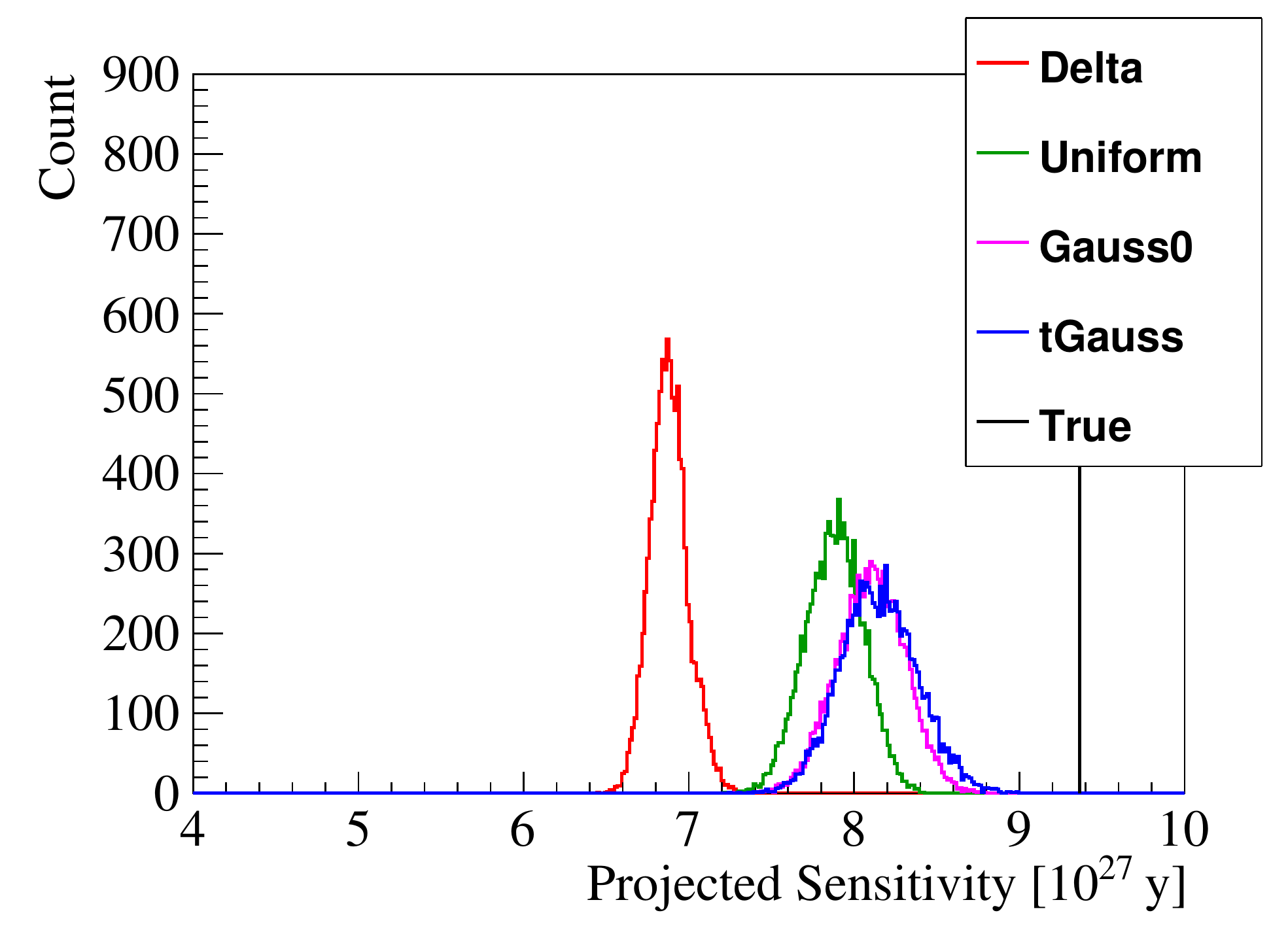} 
    \caption{$X = 10 ~\mu$Bq/kg}
    \label{fig:DeltaPriors-N20-R1}
  \end{subfigure}
  \caption{Projected sensitivities for different priors and different impurity concentrations $X$, 
for $N = 20$ and $\varepsilon = 1.585\times10^{-5}$.
Here, the distributions are visibly closer to a Gaussian shape due to 
the independence of the assay measurements of the parts.
 (Colors online)
}
  \label{fig:DeltaPriors-N20}
\end{figure}

When the impurity concentrations are close to or even below the detection limit of the HPGe assay detector,
the prior choices for upper limits become more relevant as 
upper limits are now reported more often.
Figures \ref{fig:DeltaPriors-N1} and \ref{fig:DeltaPriors-N20} show
the projected sensitivities for this detector 
with impurity concentrations $X$ lower than 100 $\mu$Bq/kg.
The distributions for different priors are seen to show some distinctive shapes as $X$ becomes smaller.

As seen in the figures, the sensitivity calculated using the Dirac delta prior
begins to visibly diverge from the \textit{true} sensitivity when $X = 40~\mu$Bq/kg, 
while the other priors only begin to do so at $X = 30~\mu$Bq/kg. 
At $X =10~\mu$Bq/kg, all distributions deviate substantially from the \textit{true} value
and they underestimate more often than overestimate.

We checked that when the hit efficiency is doubled to $\varepsilon = 6.342\times10^{-4}/N$,
almost identical features are seen at the same $X$ values. 
(This means that the \textit{true} total background rate is 2 counts per year when $X$ = 100 $\mu$Bq/kg.)
This is expected as the shapes of the distributions of the projected sensitivities 
should only depend on the assay measurements and the way they are interpreted as priors,
but not on the total background rate.

\subsection{Realistic detectors}
\label{sec:realistic}

\begin{figure}[htbp]
  \begin{multicols}{2}
  \begin{subfigure}[t]{.5\textwidth}
    \includegraphics[width=\textwidth]{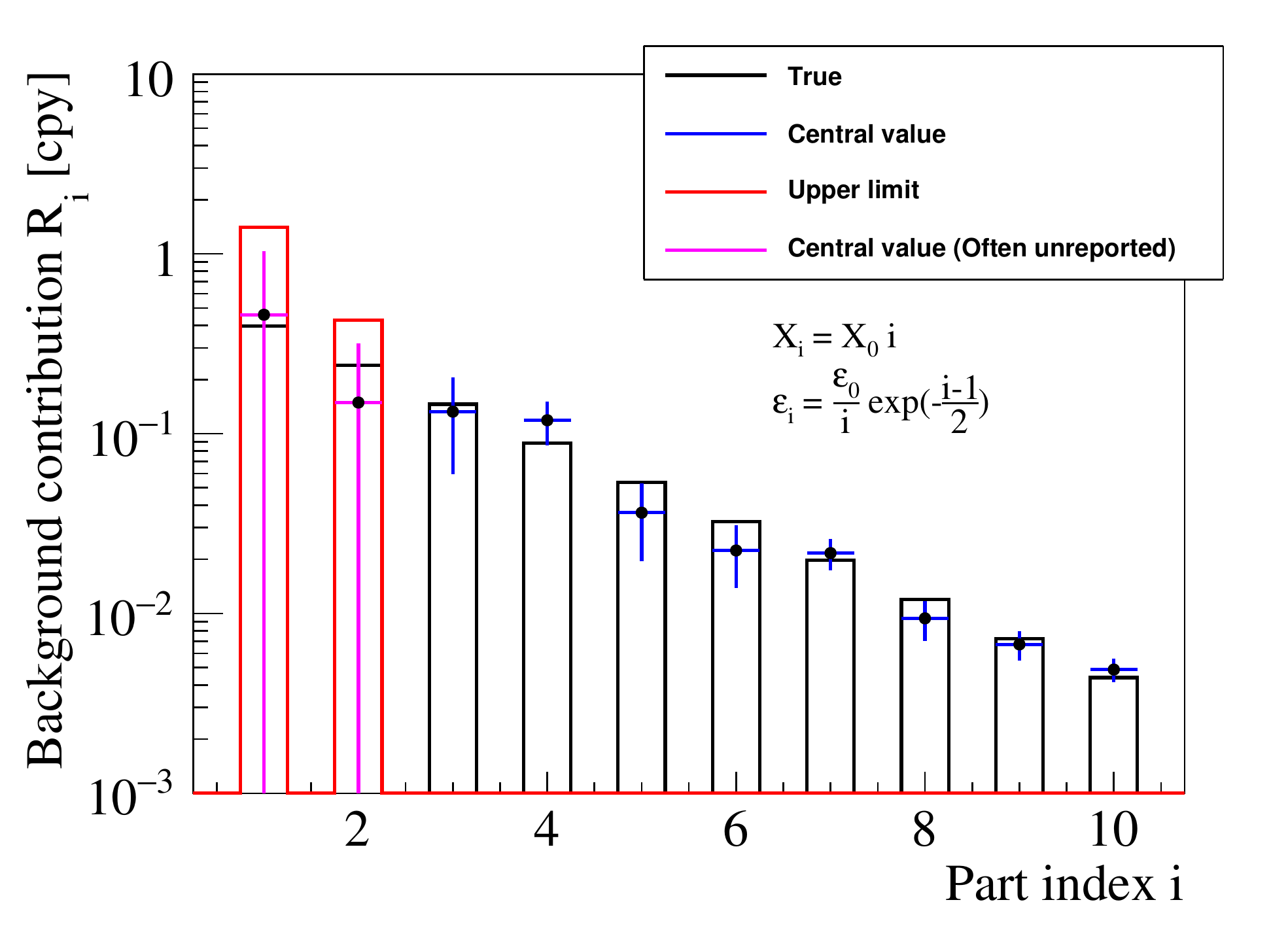}
    \caption{$X_0 = 10~\mathrm{\mu Bq/kg}$, $X_{i} = X_0 i$}
    \label{fig:realisticN-1X100}
  \end{subfigure}
  \begin{subfigure}[t]{.5\textwidth}
    \includegraphics[width=\textwidth]{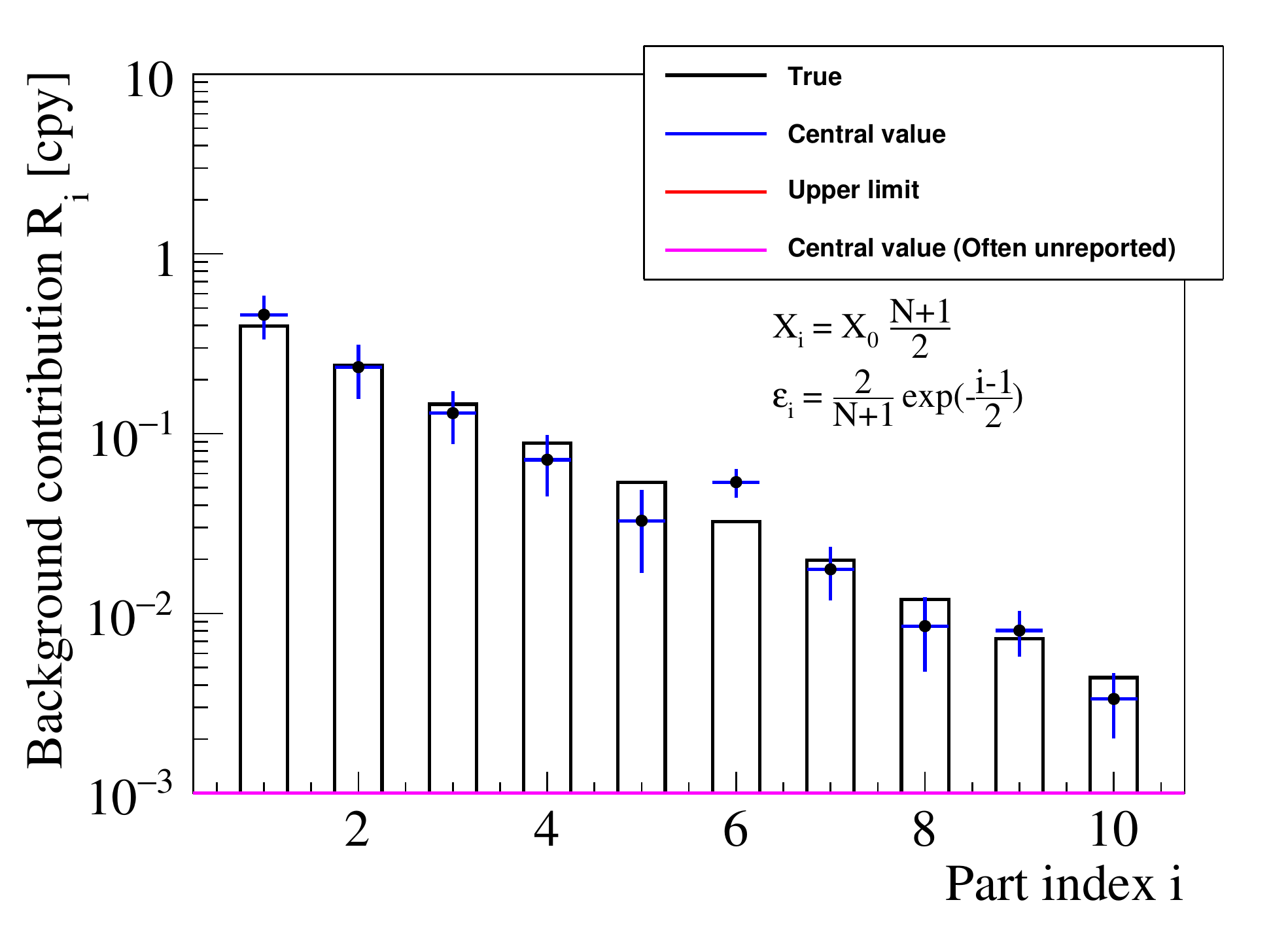}
    \caption{$X_0 = 10~\mathrm{\mu Bq/kg}$, $X_{i} = X_0 \frac{N+1}{2} = 55 ~\mathrm{\mu Bq/kg}$}
    \label{fig:realisticN-3X100}
  \end{subfigure}
  \begin{subfigure}[t]{.5\textwidth}
    \includegraphics[width=\textwidth]{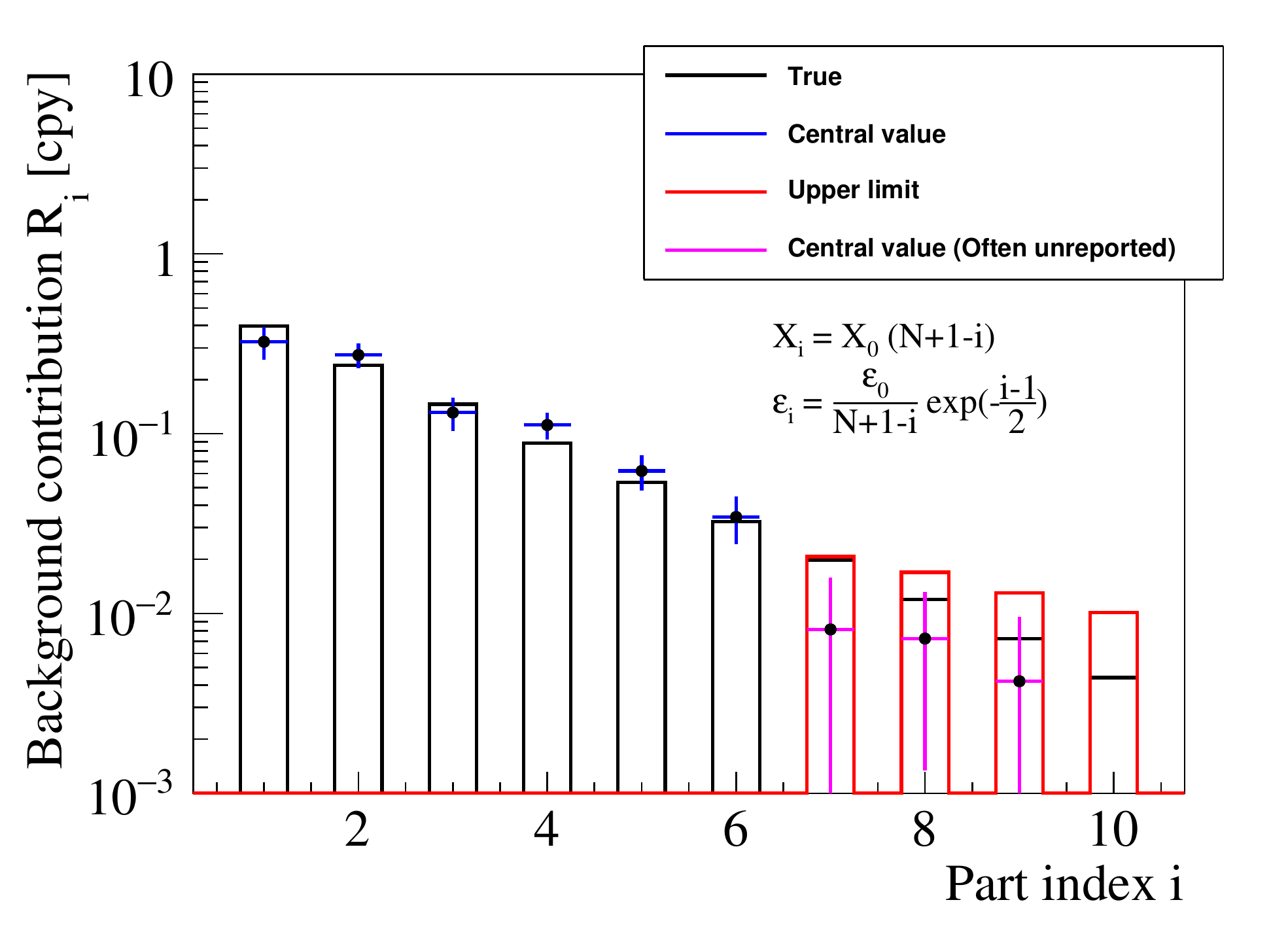}
    \caption{$X_0 = 10~\mathrm{\mu Bq/kg}$, $X_{i} = X_0 (N+1-i)$}
    \label{fig:realisticN-2X100}
  \end{subfigure}
  \begin{subfigure}[t]{.5\textwidth}
    \includegraphics[width=\textwidth]{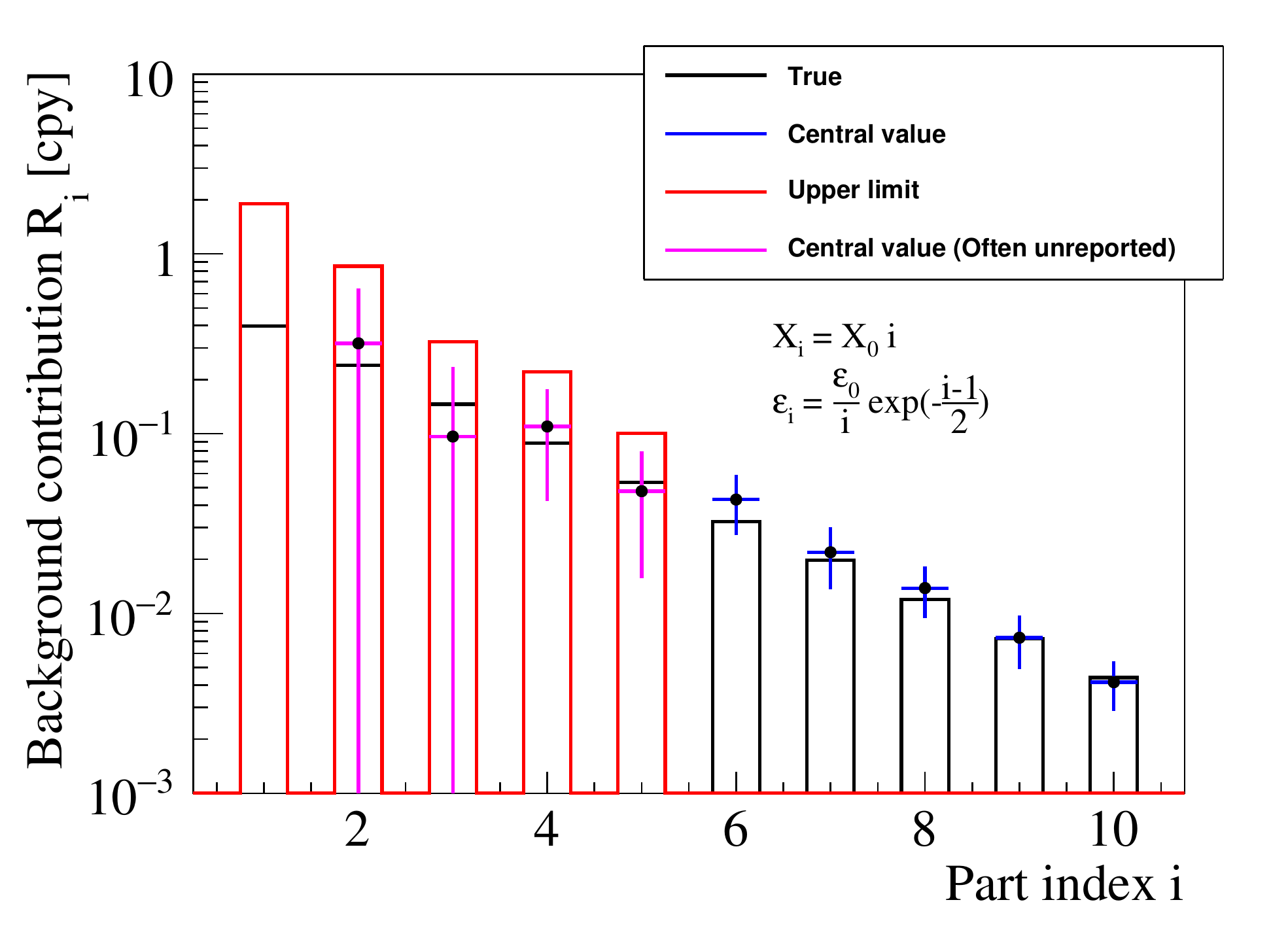}
    \caption{$X_0 = 5~\mathrm{\mu Bq/kg}$, $X_{i} = X_0 i$}
    \label{fig:realisticN-1X50}
  \end{subfigure}
  \begin{subfigure}[t]{.5\textwidth}
    \includegraphics[width=\textwidth]{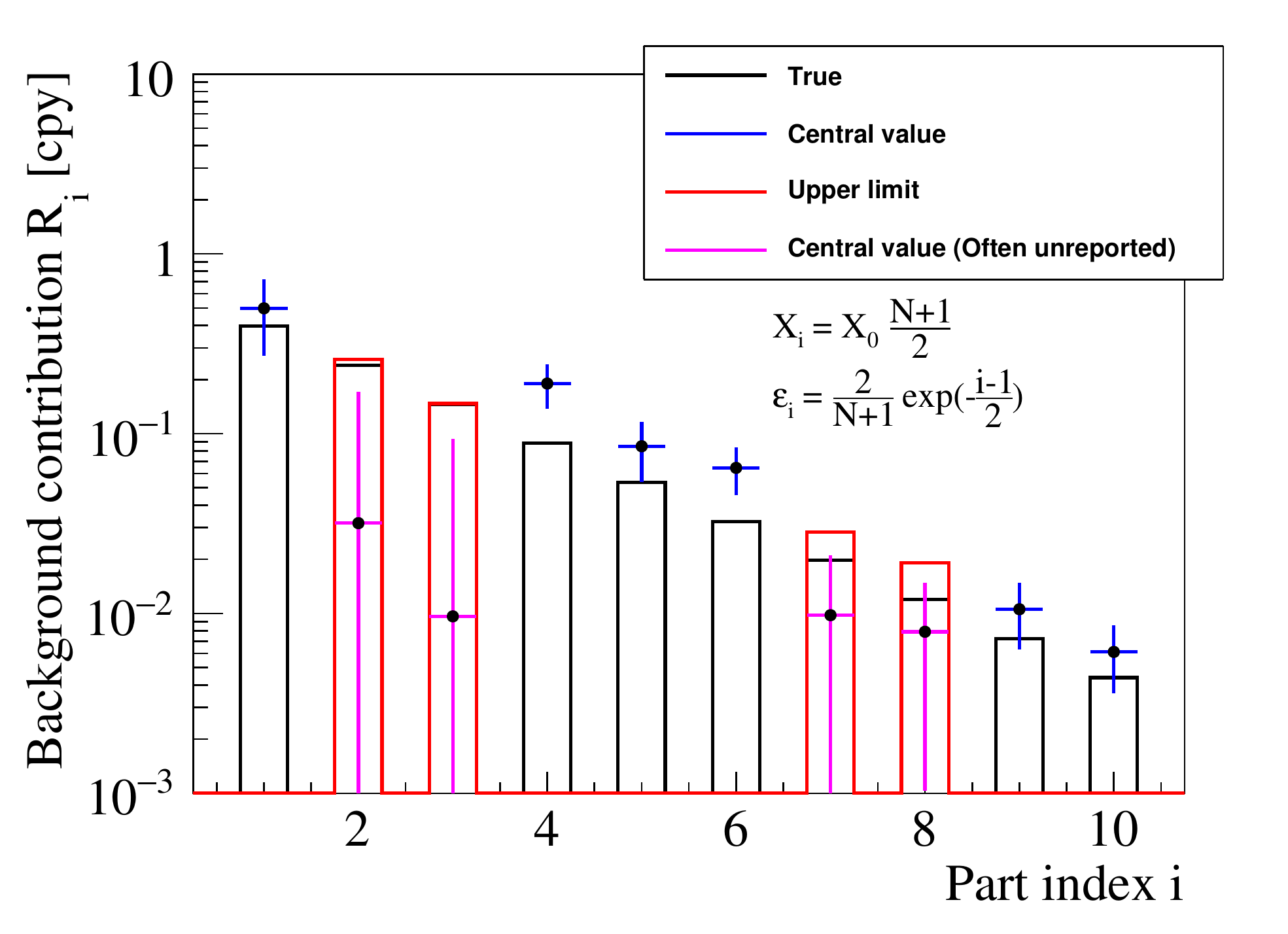}
    \caption{$X_0 = 5~\mathrm{\mu Bq/kg}$, $X_{i} = X_0 \frac{N+1}{2} = 27.5 ~\mathrm{\mu Bq/kg}$}
    \label{fig:realisticN-3X50}
  \end{subfigure}
  \begin{subfigure}[t]{.5\textwidth}
    \includegraphics[width=\textwidth]{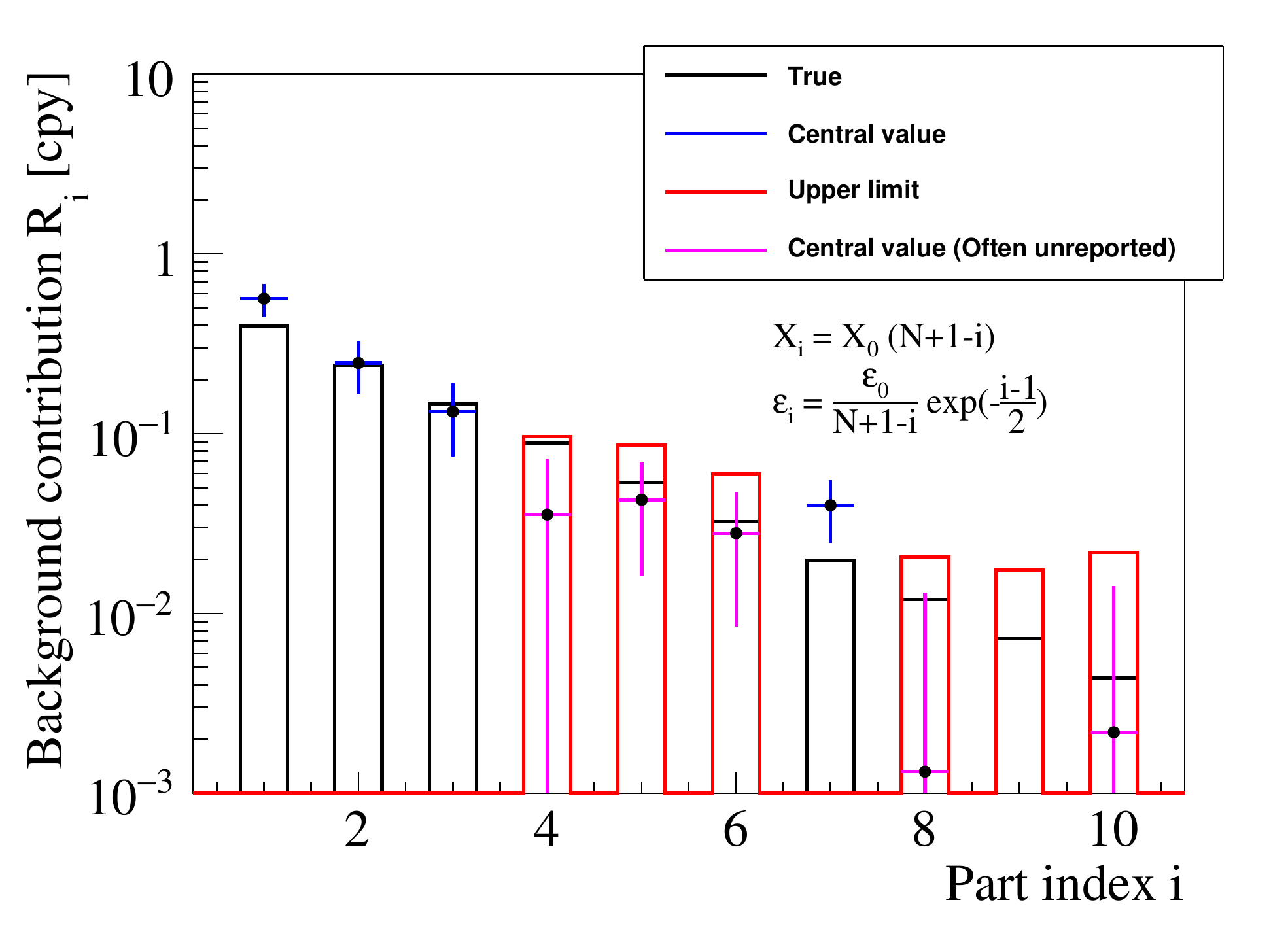}
    \caption{$X_0 = 5~\mathrm{\mu Bq/kg}$, $X_{i} = X_0 (N+1-i)$}
    \label{fig:realisticN-2X50}
  \end{subfigure}
  \end{multicols}
\caption{
The black bars show the background contributions by each part 
assuming the realistic scenarios described in Section \ref{sec:realistic}.
The colored data points and bars represent the expected background contribution based on 
the results of one realization of an assay campaign:
shown in blue are central values and errors, red upper limits,
and magenta the underlying central values and errors when an upper limit should be reported.
The distributions of projected sensitivities calculated from 
ensembles of assay campaign realizations such as this one are shown in Figure \ref{fig:realistic-sens}.
($N=10$, $\varepsilon_0 = 1.256\times10^{-3}$)
 (Colors online)
}
\label{fig:realistic}
\end{figure}

\begin{figure}[htbp]
  \begin{multicols}{2}
  \begin{subfigure}[t]{.5\textwidth}
    \includegraphics[width=\textwidth]{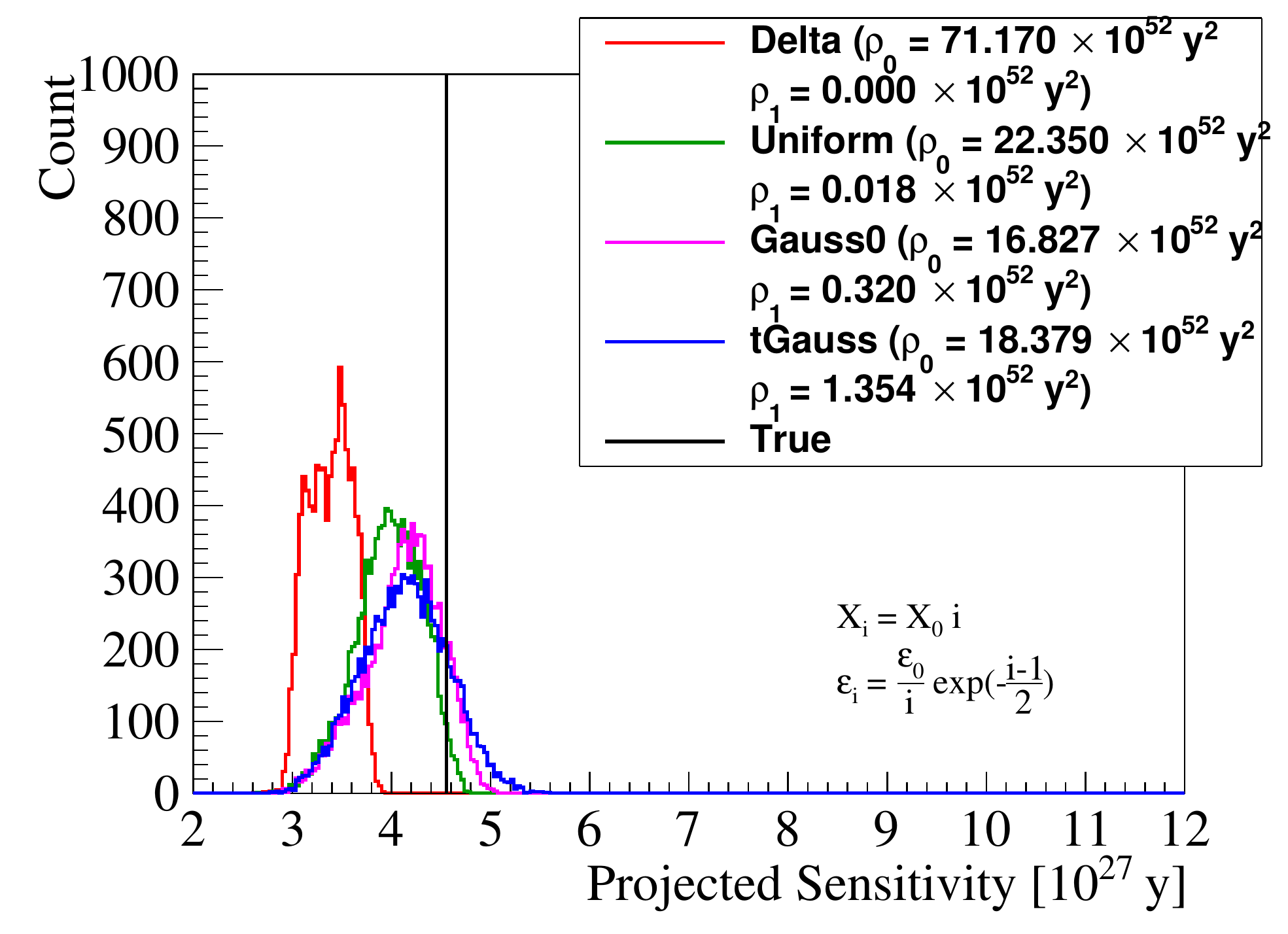}
    \caption{$X_0 = 10~\mathrm{\mu Bq/kg}$, $X_{i} = X_0 i$}
    \label{fig:realisticN-1X100-sens}
  \end{subfigure}
  \begin{subfigure}[t]{.5\textwidth}
    \includegraphics[width=\textwidth]{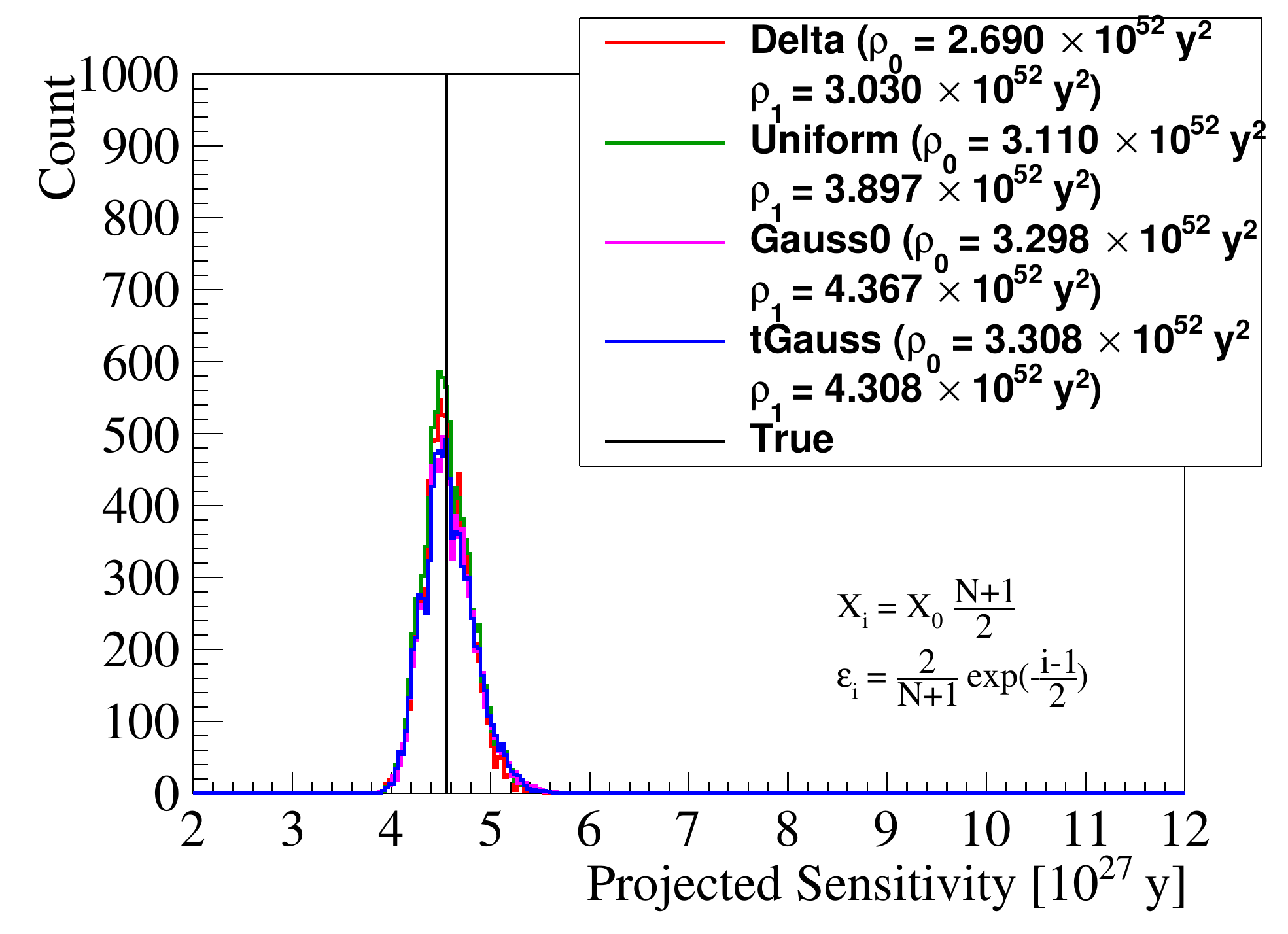}
    \caption{$X_0 = 10~\mathrm{\mu Bq/kg}$, $X_{i} = X_0 \frac{N+1}{2} = 55 ~\mathrm{\mu Bq/kg}$}
    \label{fig:realisticN-3X100-sens}
  \end{subfigure}
  \begin{subfigure}[t]{.5\textwidth}
    \includegraphics[width=\textwidth]{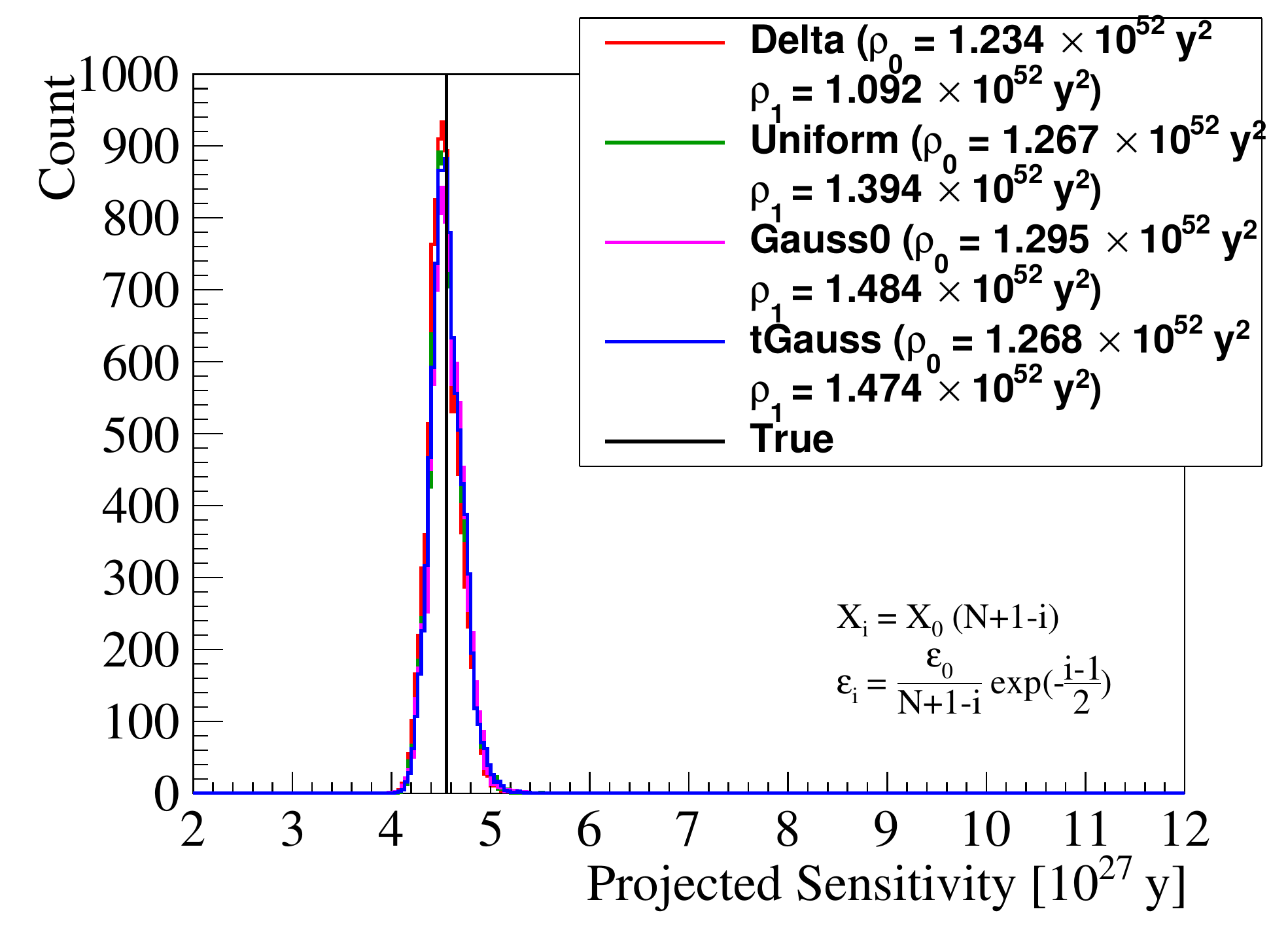}
    \caption{$X_0 = 10~\mathrm{\mu Bq/kg}$, $X_{i} = X_0 (N+1-i)$}
    \label{fig:realisticN-2X100-sens}
  \end{subfigure}
  \begin{subfigure}[t]{.5\textwidth}
    \includegraphics[width=\textwidth]{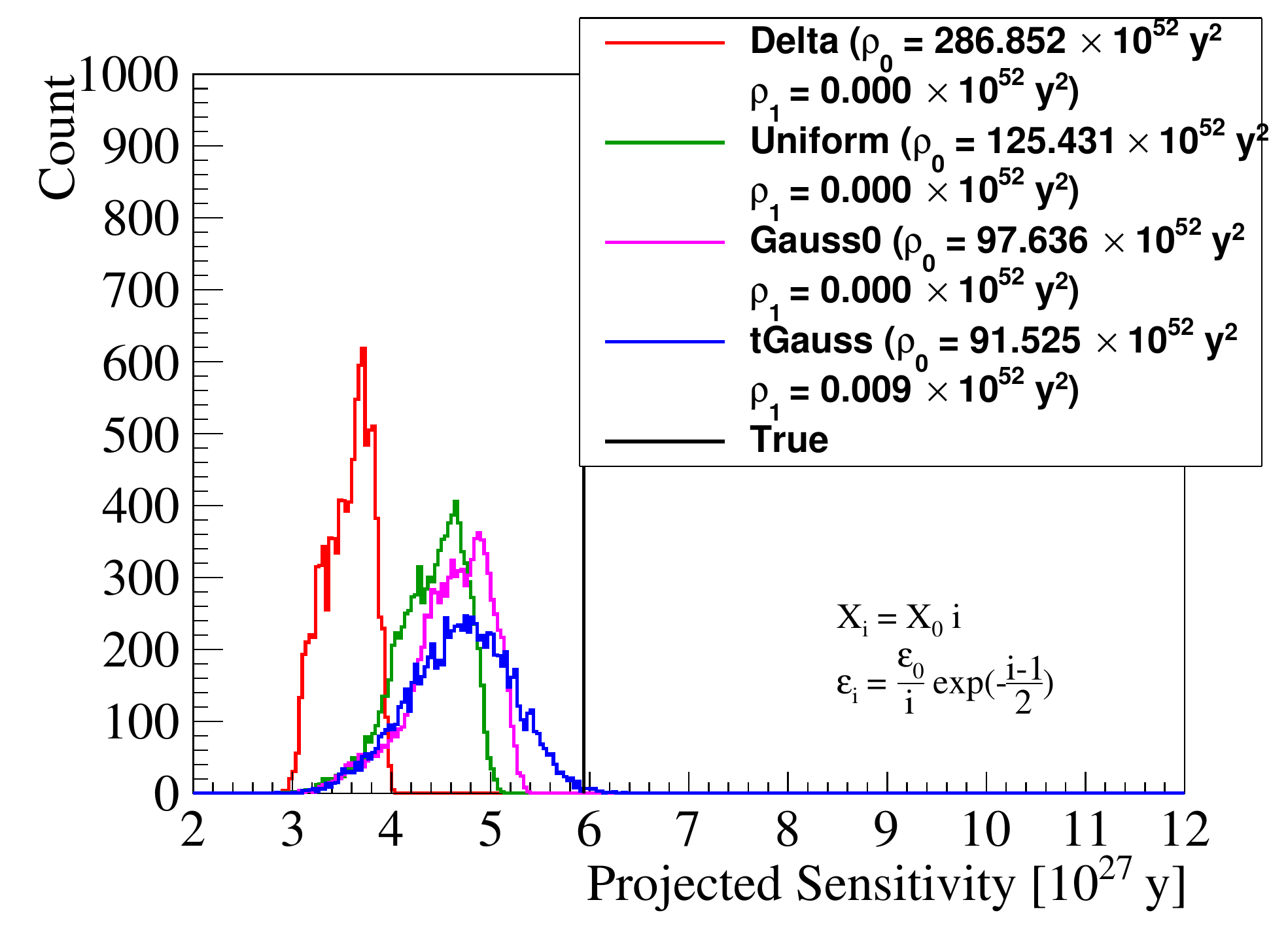}
    \caption{$X_0 = 5~\mathrm{\mu Bq/kg}$, $X_{i} = X_0 i$}
    \label{fig:realisticN-1X50-sens}
  \end{subfigure}
  \begin{subfigure}[t]{.5\textwidth}
    \includegraphics[width=\textwidth]{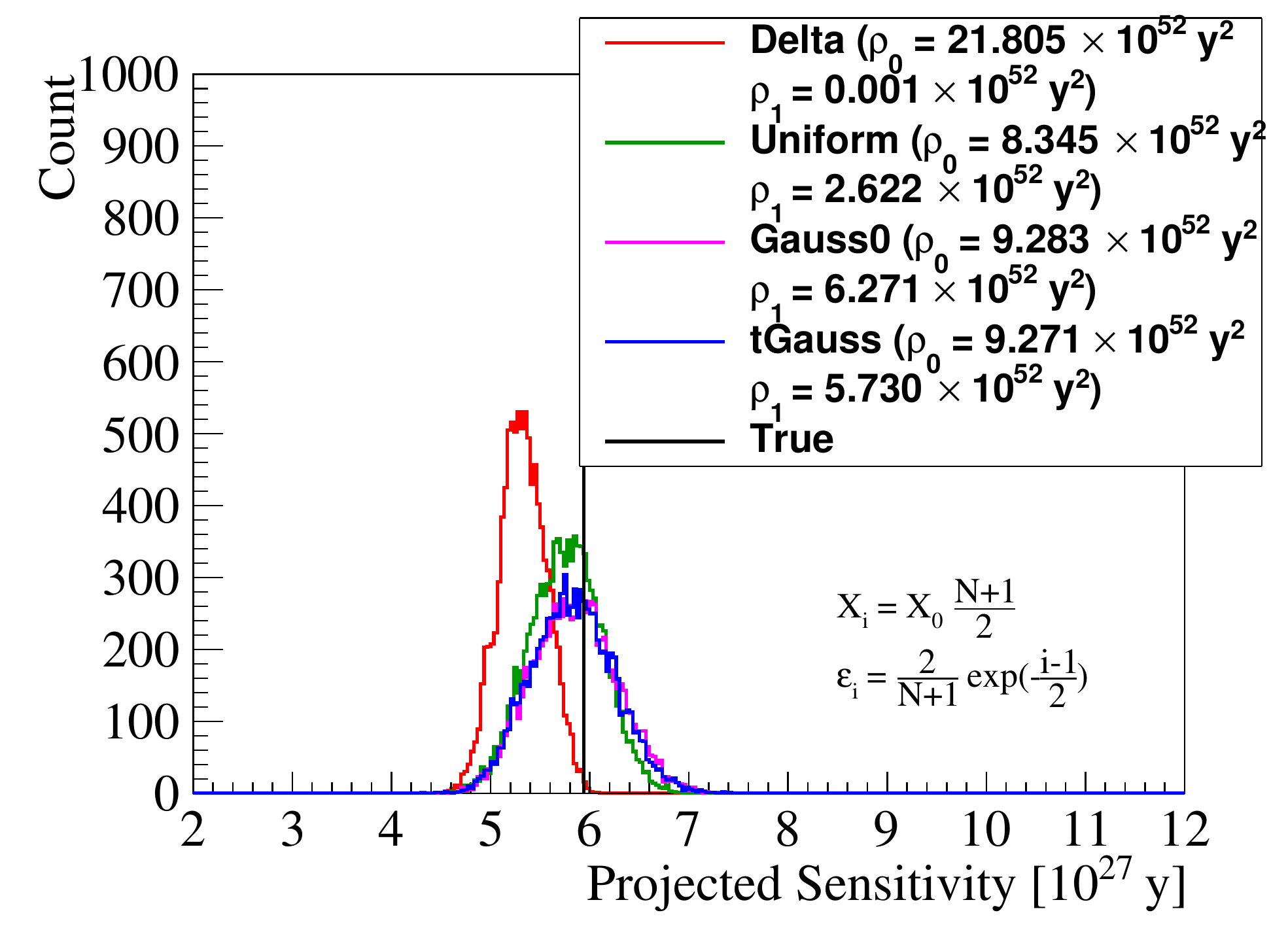}
    \caption{$X_0 = 5~\mathrm{\mu Bq/kg}$, $X_{i} = X_0 \frac{N+1}{2} = 27.5 ~\mathrm{\mu Bq/kg}$}
    \label{fig:realisticN-3X50-sens}
  \end{subfigure}
  \begin{subfigure}[t]{.5\textwidth}
    \includegraphics[width=\textwidth]{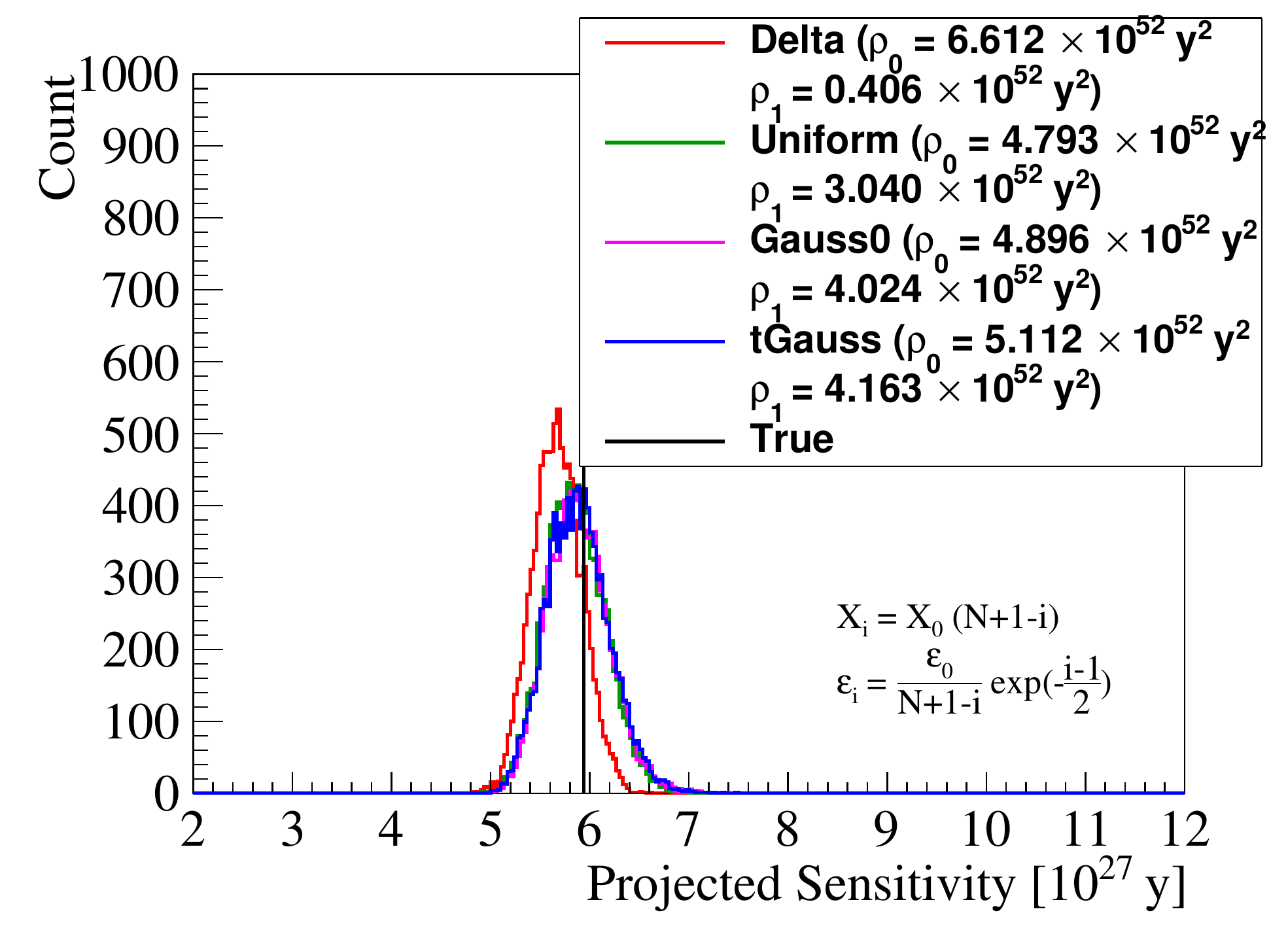}
    \caption{$X_0 = 5~\mathrm{\mu Bq/kg}$, $X_{i} = X_0 (N+1-i)$}
    \label{fig:realisticN-2X50-sens}
  \end{subfigure}
  \end{multicols}
\caption{
Projected sensitivities for the realistic scenarios described in Section \ref{sec:realistic}.
Each data point in the histograms is calculated using the assay results 
from a realization of an assay campaign
such as the one shown in Figure \ref{fig:realistic}. 
($N=10$, $\varepsilon_0 = 1.256\times10^{-3}$)
 (Colors online)
}
\label{fig:realistic-sens}
\end{figure}

So far, we have only considered detectors with identical parts, which contribute equally to the total background.
As we have observed in general, 
in an actual detector, 
the background contributions by parts often loosely follow an exponential distribution 
(e.g. \cite{CuorePrior,NexoPrior})
and so will we stipulate for this analysis.

Let us consider a detector that has 10 parts (N=10) and that 
the background contribution by the $i$th part ($R_i$) is proportional to $e^{-\frac{i-1}{2}}$.
In other words, part 1 contributes most to the total background and part 10 the least,
and their contributions differ by about a factor of 100.

Let us consider three scenarios
for the impurity concentrations of each part, 
\begin{itemize}
\item Scenario 1: $X_i$ increases linearly with $i$: $X_i = X_0 i$. 
\item Scenario 2: $X_i$ remains constant: $X_i = X_0 \frac{N+1}{2}$.
\item Scenario 3: $X_i$ decreases linearly with $i$: \\$X_i = X_0 (N+1-i)$. 
\end{itemize}
By fixing the total background rate $\sum R_i$, the hit efficiencies for each part can be defined.

Figures \ref{fig:realisticN-1X100}, \ref{fig:realisticN-3X100}, and \ref{fig:realisticN-2X100} illustrate the 
three scenarios when $X_0$ is set to 10 $\mu$Bq/kg, and 
Figures \ref{fig:realisticN-1X50}, \ref{fig:realisticN-3X50}, and \ref{fig:realisticN-2X50} 
show similar plots for $X_0$ = 5 $\mu$Bq/kg.
Figure \ref{fig:realistic-sens} shows the distributions of the projection sensitivities for the above scenarios.

With a configuration such as Scenario 1, 
the sensitivity estimation is susceptible to the prior choice for upper limits,
as the background contribution from the largest contributor is more likely derived from an upper limit.
Whereas in Scenario 3, prior choice for central values plays a more important role in sensitivity estimation.

\section{Discussion}
\label{sec:Discussion}

In the previous section, 
we presented the raw results of calculating projected sensitivities 
under different experimental scenarios and choices of priors used to interpret assay results. 
Now we proceed to evaluate these results.

\subsection{Quantifying conservativeness} 
\begin{figure}[htbp]
\includegraphics[width=\textwidth]{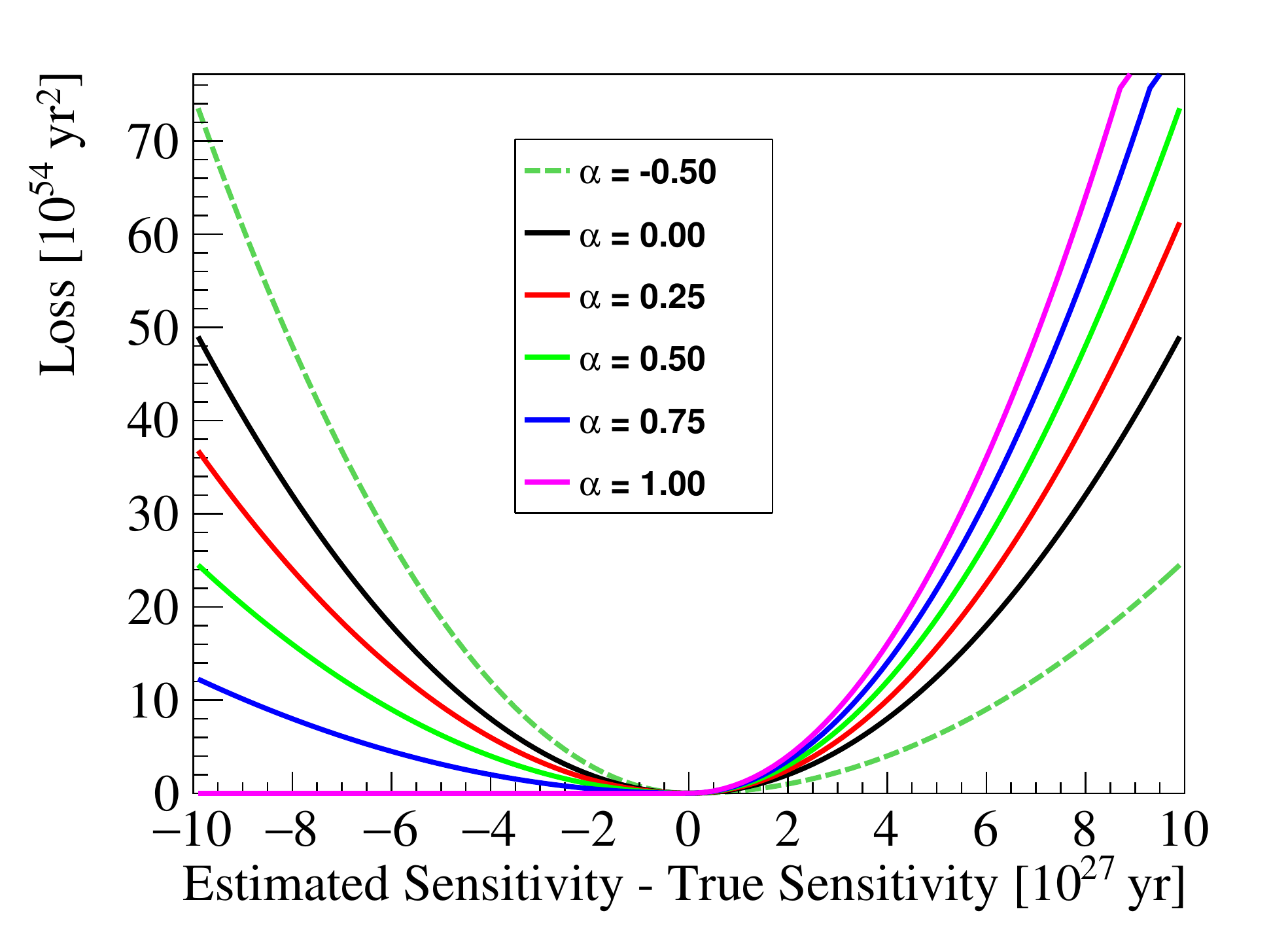}  
\caption{Examples of the loss function defined in Equation \ref{eq:loss} with different values of $\alpha$. 
In this paper, only positive values of $\alpha$ are considered.
 (Colors online)
}
\label{fig:alphaloss}
\end{figure}

One of the factors in evaluating a choice of assay result prior is 
the perceived \textit{conservativeness} of the resulting projected sensitivity.
Necessarily then, it is important to quantify \textit{conservativeness} in order to guide our choice.
To our knowledge this step of quantifying a notion of conservativeness
has not been considered in the literature of low background physics experiments.
We could consider projected sensitivity as an estimator of the \textit{true} sensitivity
where the prior choice can be seen as a parameter for this estimator. 
Choosing a different prior for assay results will generate a different estimator.
To evaluate the goodness of an estimator, the quadratic loss function (QLF) is often used.
However, being symmetric, QLF is inadequate in our situation because,
in the case of limit-setting experiments, 
we may disfavor an overestimate more than an underestimate
while QLF cannot reflect that. 
To allow assigning different losses to
underestimates and overestimates,
we can define an asymmetric loss function as follows, 

\begin{equation}
L_\alpha(X,\hat{S}_p(Y)) = \begin{cases} 
\frac{1+\alpha}{2} [\hat{S}_p(Y)-S(X)]^2 & \textrm{if}~\hat{S} > S \\
\frac{1-\alpha}{2} [\hat{S}_p(Y)-S(X)]^2 & \textrm{if}~\hat{S} \le S \\
\end{cases}
\label{eq:loss}
\end{equation}
where 
$X$ denotes the set of \textit{true} impurity concentrations of all parts $\{X_i\}$
and $Y$ denotes the set of 
assay results $\{Y_i\}$ where $Y_i$ = $\mu_i\pm\sigma_I$ or $\ell_i$.
$\hat{S}_p = \hat{S}_p(Y)$ is the projected sensitivity 
calculated from assay results $Y$ using the prior choice $p$.
We propose to use $\alpha$ as a measure of conservativeness in this study.
A positive $\alpha$ represents more disfavor for overestimates than underestimates, 
a negative $\alpha$ has the opposite effect, 
and $\alpha = 0$ is effectively the ordinary QLF which treats underestimates and overestimates equally. 
Figure \ref{fig:alphaloss} shows some example loss functions at different $\alpha$'s.

An estimator is optimal if it minimizes 
the expected value of loss as a function of the \textit{true} impurity concentrations $X$,
also known as \textit{risk},
which is defined as,

\begin{equation}
\rho_\alpha(X,\hat{S}_p) = \int L_\alpha(X,\hat{S}_p(Y)) \prod_i A_{X_i}(Y_i) dY_i 
\end{equation}
where
$A_{X_i}(Y_i)$ is the probability distribution of $Y_i$ reported by an assay measurement given the \textit{true} impurity concentration $X_i$.
(Notice that $\rho_0$ is the usual mean squared error.)
We will use \textit{risk} ($\rho_{\alpha}$) defined above as the figure of merit to compare different prior choices.

\subsubsection{Prior choice for central values}

If we take the risks for the Dirac delta prior and the Gaussian prior at face value
(as shown in the legends in Figure \ref{fig:deltavsgauss}),
the Dirac delta prior seems to be the preferred choice whether $\alpha$ is 0 or 1.
In fact, since $\rho_\alpha$ depends linearly on $\alpha$, 
we can deduce that, $\rho_\alpha(X,\hat{S}_{D}) < \rho_\alpha(X,\hat{S}_{G})$ for all $\alpha \in [0,1]$.
Therefore, the Dirac delta prior is preferred regardless of your choice of $\alpha \in [0,1]$.
However, the advantage diminishes as $N$ increases.

As compared to the differences due to the prior choice for upper limits, 
the difference between the two priors for central values appears small.
Therefore, the Dirac delta prior is preferred but not strongly.


\subsubsection{Prior choice for upper limits}
As indicated in Figure \ref{fig:realistic-sens},
the best prior choice for upper limits,
as suggested by the figure-of-merit $\rho_\alpha$, 
appears to depend on the preferred value of $\alpha$.
For those who prefer $\alpha$ = 1 (most conservative),
the Dirac delta prior is clearly the best choice among the four priors considered.
For $\alpha$ = 0, the situation is more complex. 
For the two cases where central values are more often reported
(Figures \ref{fig:realisticN-3X100-sens} and \ref{fig:realisticN-2X100-sens}), 
the Dirac delta prior is the best choice. 
For the four other cases where upper limits are more often reported, there is no clear best choice among
the Uniform, the Gaussian-at-zero, and the Truncated Gaussian priors.
It seems to depend on the particular distribution of impurity concentrations,
though these three priors are clearly better than the Dirac delta prior in these four cases.

\subsection{Appropriate values of $\alpha$}

\begin{figure}[htbp]
  \begin{subfigure}[t]{.5\textwidth}
    \includegraphics[width=\textwidth]{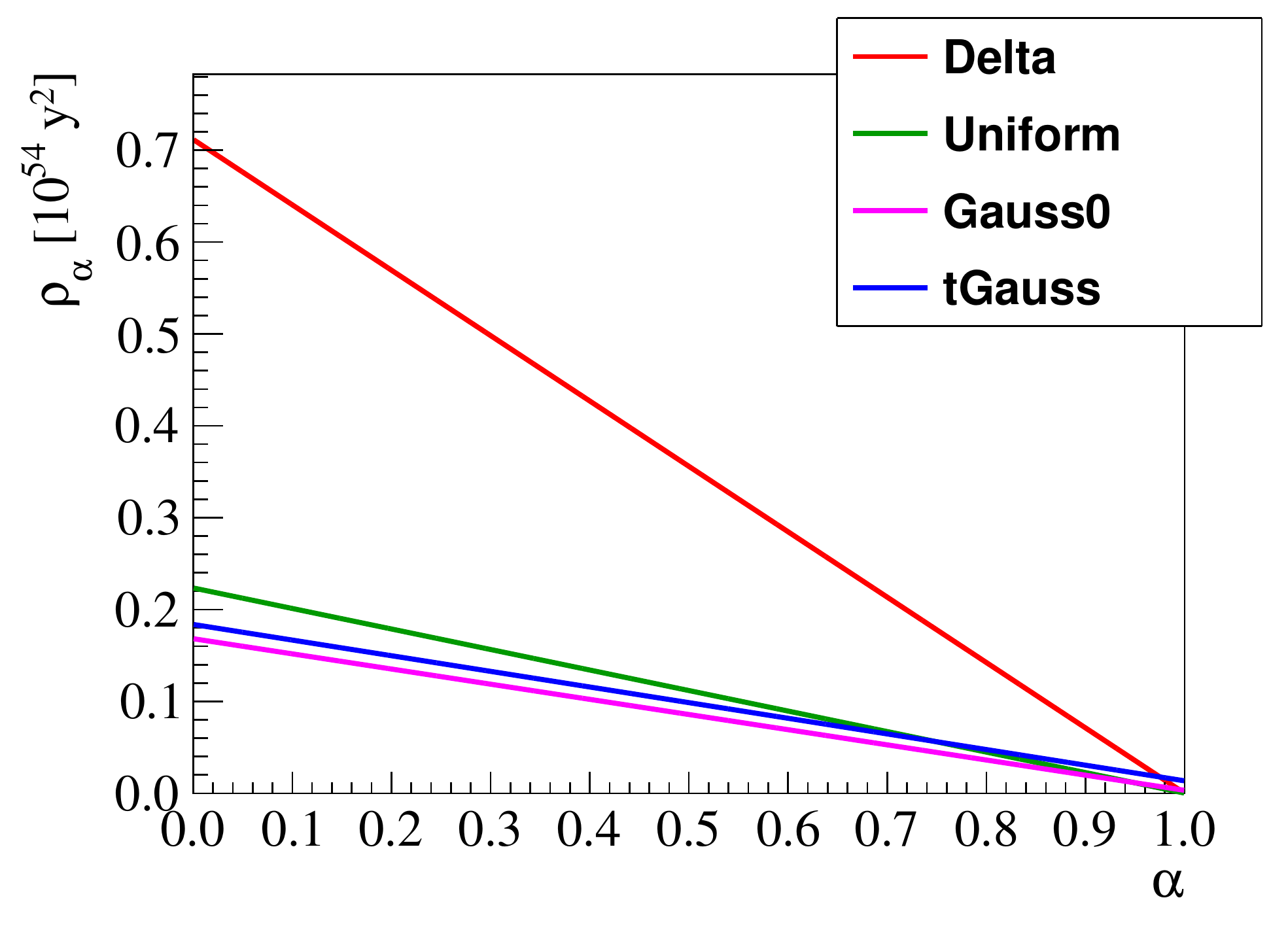}
    \caption{$X_0 = 10~\mathrm{\mu Bq/kg}$, $X_{i} = X_0 i$}
    \label{fig:sumriskN-1X100-sens}
  \end{subfigure}
  \begin{subfigure}[t]{.5\textwidth}
    \includegraphics[width=\textwidth]{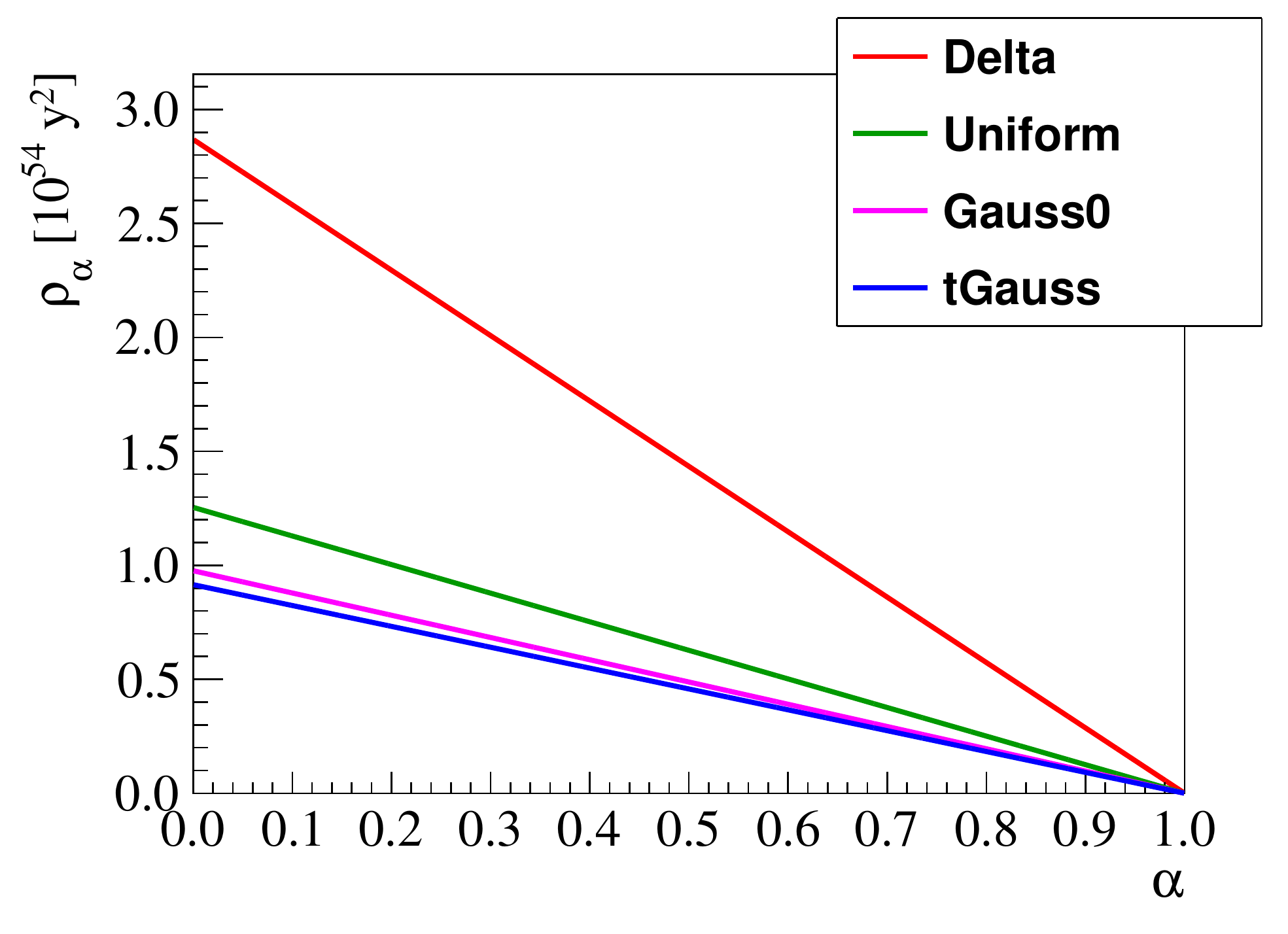}
    \caption{$X_0 = 5~\mathrm{\mu Bq/kg}$, $X_{i} = X_0 i$}
    \label{fig:sumriskN-1X50-sens}
  \end{subfigure}
\caption{Risk as a function of $\alpha$ for Scenario 1 considered in Section \ref{sec:realistic}.
Low risk (vertical axis) may be considered subjectively ``better''.
 (Colors online)
}
\end{figure}

As shown in the results above, risk depends on the value of $\alpha$.
Now, a question remains:
How to assign a value to $\alpha$ 
that reflects our relative disfavor for
underestimates and overestimates?
Here are some considerations:
\begin{itemize}
\item
$\alpha$ is a subjectively-chosen ``hidden'' variable 
which manifests itself only through the choice of a prior. 
Therefore, values of $\alpha$ that result in the same prior choice 
can be considered identical for the purpose of this study.
\item
In Scenario 1 considered above where upper limits dominate, 
the Dirac delta prior is \textit{better} than the other priors
only when $\alpha$ is very close to 1 
($>\sim$0.98, see Figures \ref{fig:sumriskN-1X100-sens} and \ref{fig:sumriskN-1X50-sens}).
For other values of $\alpha$, the Truncated Gaussian prior, the Gaussian-at-zero prior, and the Uniform prior
all give similar risk values, 
indicating that they are almost equally \textit{good} when judged by this metric.
\end{itemize}

\label{fig:sumrisk}

\begin{table}
\centering
\begin{tabular}{c|cc|c}
Collaboration & Central values and errors & Upper limits & Ref. \\
\hline
Majorana Demonstrator & Delta & Delta  &  \cite{ABGRALL201622} \\
NEXT & Delta & Delta  & \cite{NextPrior} \\
PandaX III & Delta & Delta  & \cite{PandaxPrior} \\
CUORE & Gaussian & \textit{(included as systematics)} & \cite{CuorePrior}  \\
CUORE-0 & Gaussian &  Half Gaussian at zero  & \cite{Cuore0Prior} \\
nEXO & Gaussian &  Half Gaussian at zero & \cite{NexoPrior} \\
\end{tabular}
\caption{Priors chosen by some experiment collaborations for their sensitivity projections.}
\label{tab:expts}
\end{table}

In practice, some collaborations effectively choose the Dirac delta prior for upper limits (See Table \ref{tab:expts}).
This reflects their implicit choice of an $\alpha$ that is possibly very close to 1.
Essentially this choice uses positively biased assay results in sensitivity calculations,
to result in a negatively biased projected sensitivity,
apparently being ``conservative'' in reporting a projected sensitivity.
Initially, this appears to be a prudent choice.
However, as seen in the projected sensitivity distributions shown previously in this paper, 
overestimations cannot be avoided no matter which prior is chosen.
Moreover,
taking the upper limits verbatim poses another problem. 
The upper limits may be reported at any confidence levels (say, 90\%, 95\% or 3$\sigma$),
depending on the common practice of the particular assay technique and/or assay facility.
They result in different amounts of negative bias in the projected sensitivities.
However, there is no \textit{a priori} reason why a negative bias of a certain magnitude should be introduced,
other than simply convenience (``just use the upper limit as reported'').
Thus, the projected sensitivity calculated is contingent on the choice of confidence level by the assayer,
but not entirely on the radioimpurity concentrations of the samples
and the assay sensitivities.

To remove this subjectivity,
the response would be 
to aim for an unbiased estimate of the sensitivity (i.e. $\alpha$ = 0).
The Truncated Gaussian prior is arguably the best choice because in addition to its low risk value at $\alpha$ = 0,
it is also motivated by Bayesian considerations on the assay measurements (as discussed in \ref{sec:bayes}.)
Since forming the Truncated Gaussian prior requires information that is sometimes unavailable 
(e.g. $\mu$ and $\sigma$ when only the upper limit is reported),
in those cases, both the Gaussian-at-zero prior and the Uniform prior are viable alternatives.

\subsection{Estimating the spread of projected sensitivity, $\sigma_S$}
The remaining concern for overestimation can be mitigated by
quantifying the spread ($\sigma_S$) of the projected sensitivity 
due to statistical uncertainty in the assay results.

To estimate $\sigma_S$, 
we can treat the assay results as Gaussian-distributed nuisance parameters,
and propagate the uncertainties using a Monte Carlo method. 
This is sometimes known as the ``pull'' method (described, for example, in \cite{PullMethod}.)
The quantity $\sigma_S$ can then be seen as 
the standard deviation of the resulting projected sensitivity distribution.

We could check to see if $\sigma_S$ represents the correct coverage probabilities for the above realistic scenarios 
using a $z$-score defined as,

\begin{equation}
z = \frac{\hat{S} - S}{\sigma_S}
\end{equation}
where $\hat{S}$ and $S$ are respectively the projected sensitivity 
(calculated using the Truncated Gaussian prior)
and the \textit{true} sensitivity.
We can then compare the distribution of $z$ with the standard Gaussian distribution.

\begin{figure}[htbp]
  \begin{multicols}{2}
  \begin{subfigure}[t]{.5\textwidth}
    \includegraphics[width=\textwidth]{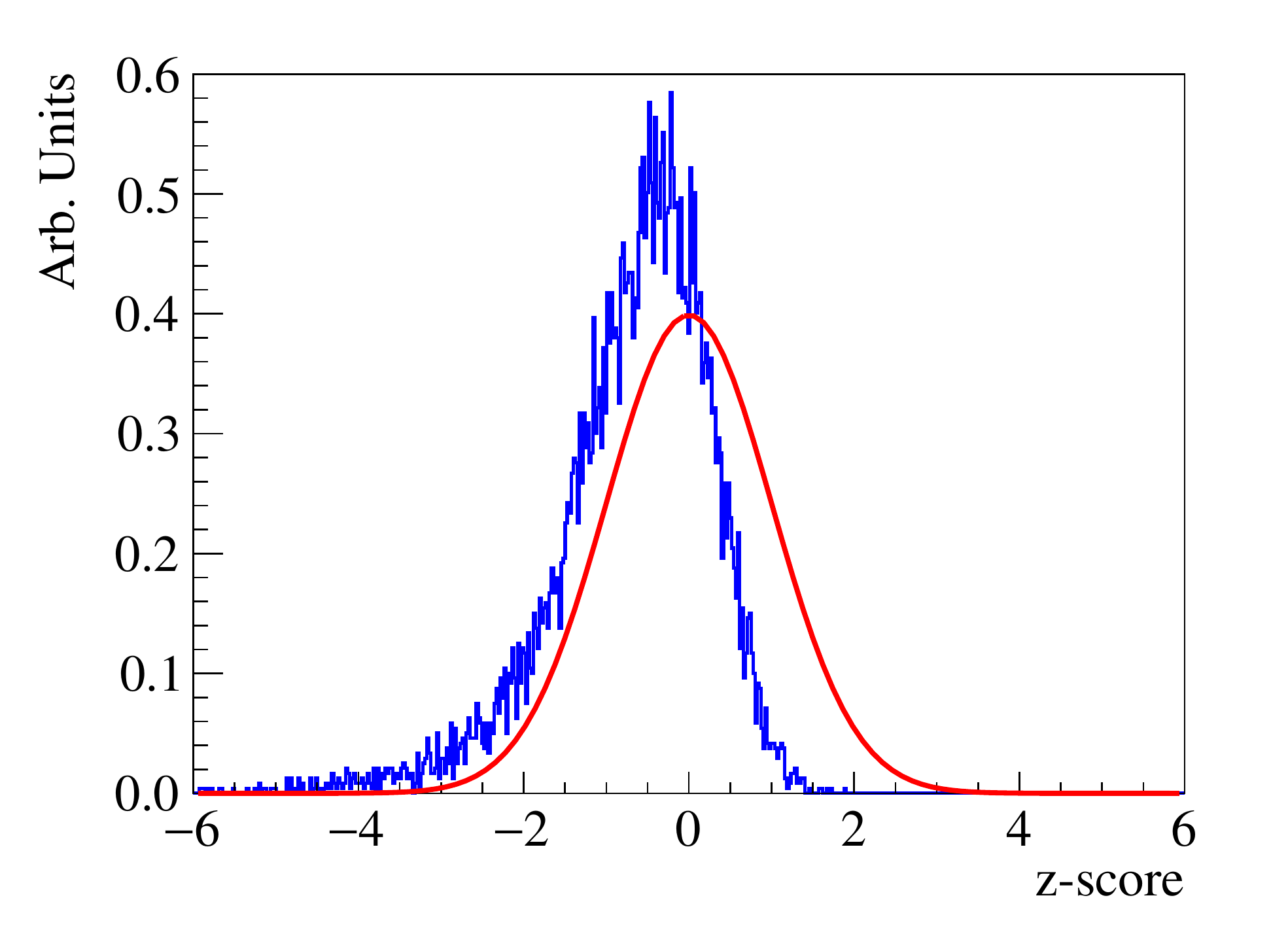}
    \caption{$X_0 = 10~\mathrm{\mu Bq/kg}$, $X_{i} = X_0 i$}
    \label{fig:zstatN-1X100}
  \end{subfigure}
  \begin{subfigure}[t]{.5\textwidth}
    \includegraphics[width=\textwidth]{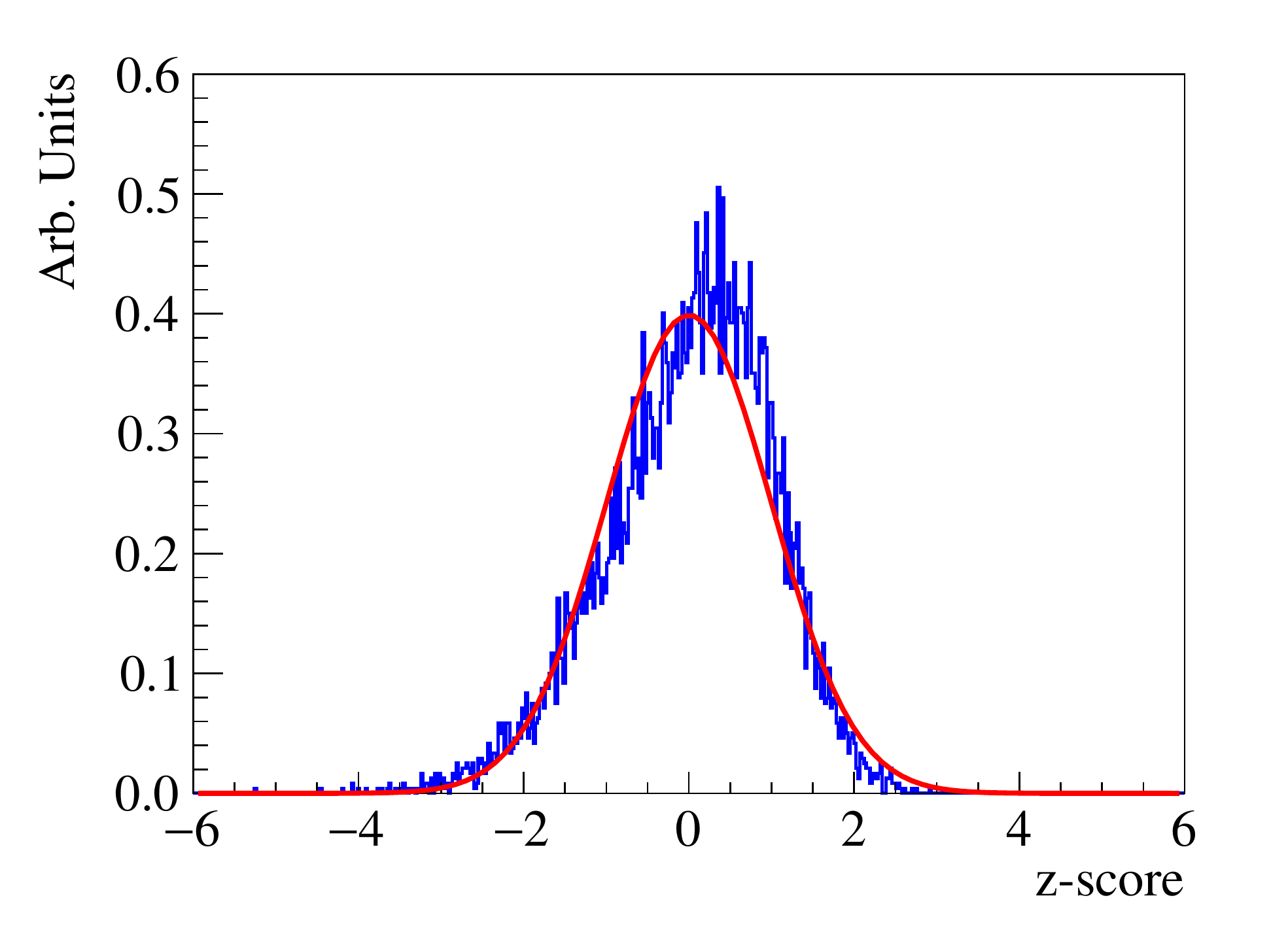}
    \caption{$X_0 = 10~\mathrm{\mu Bq/kg}$, $X_{i} = X_0 \frac{N+1}{2} = 55 ~\mathrm{\mu Bq/kg}$}
    \label{fig:zstatN-3X100}
  \end{subfigure}
  \begin{subfigure}[t]{.5\textwidth}
    \includegraphics[width=\textwidth]{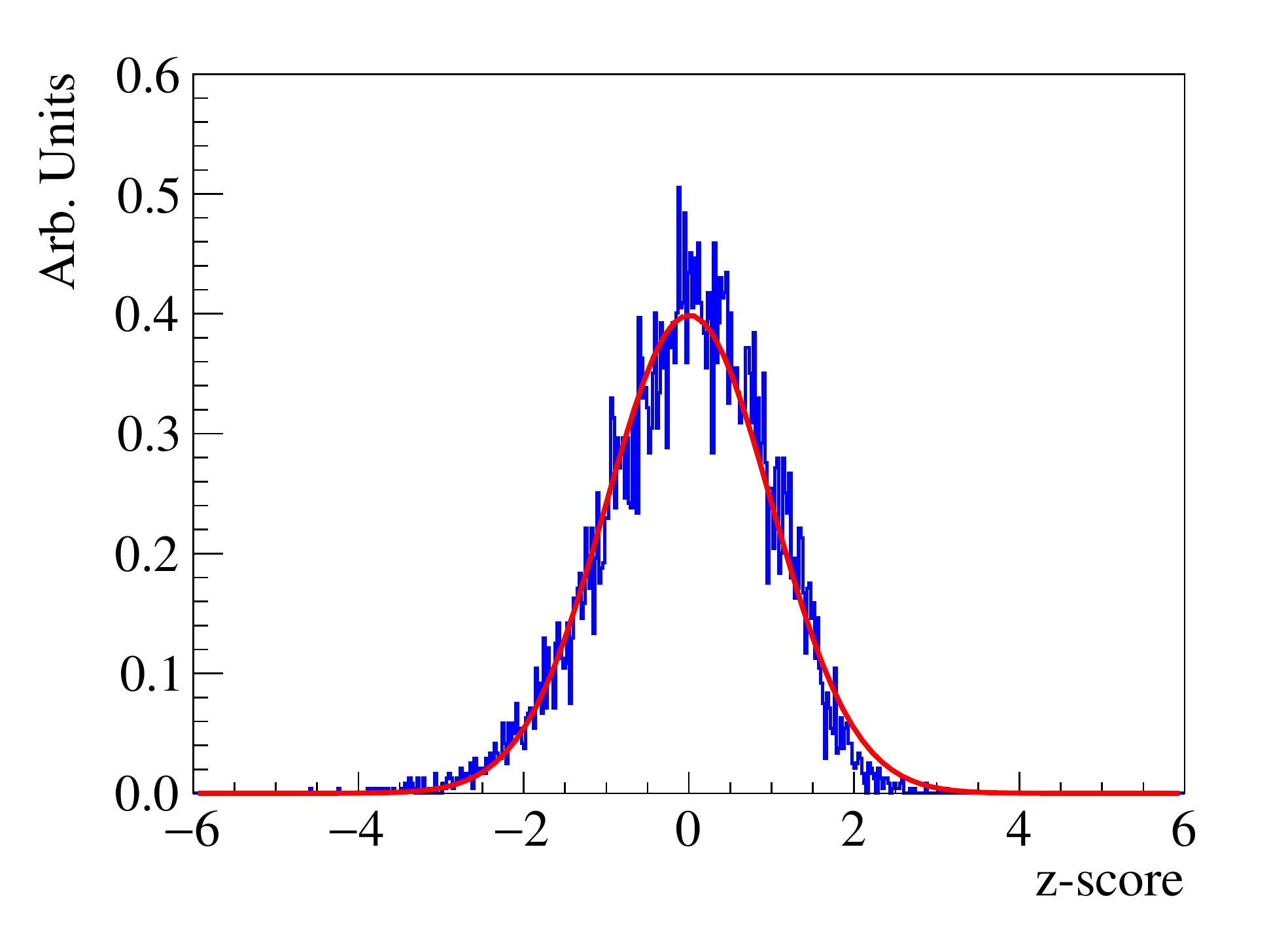}
    \caption{$X_0 = 10~\mathrm{\mu Bq/kg}$, $X_{i} = X_0 (N+1-i)$}
    \label{fig:zstatN-2X100}
  \end{subfigure}
  \begin{subfigure}[t]{.5\textwidth}
    \includegraphics[width=\textwidth]{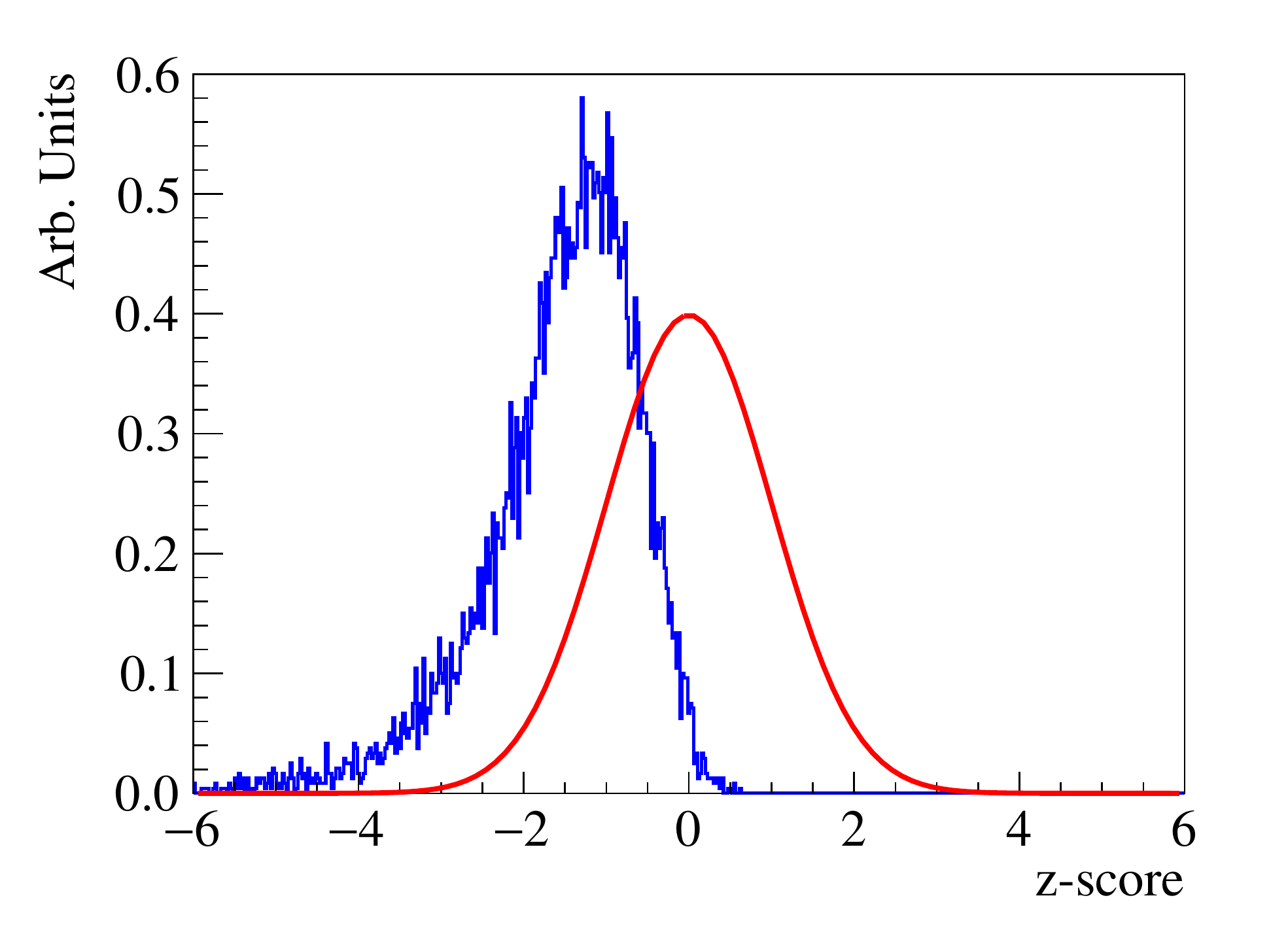}
    \caption{$X_0 = 5~\mathrm{\mu Bq/kg}$, $X_{i} = X_0 i$}
    \label{fig:zstatN-1X50}
  \end{subfigure}
  \begin{subfigure}[t]{.5\textwidth}
    \includegraphics[width=\textwidth]{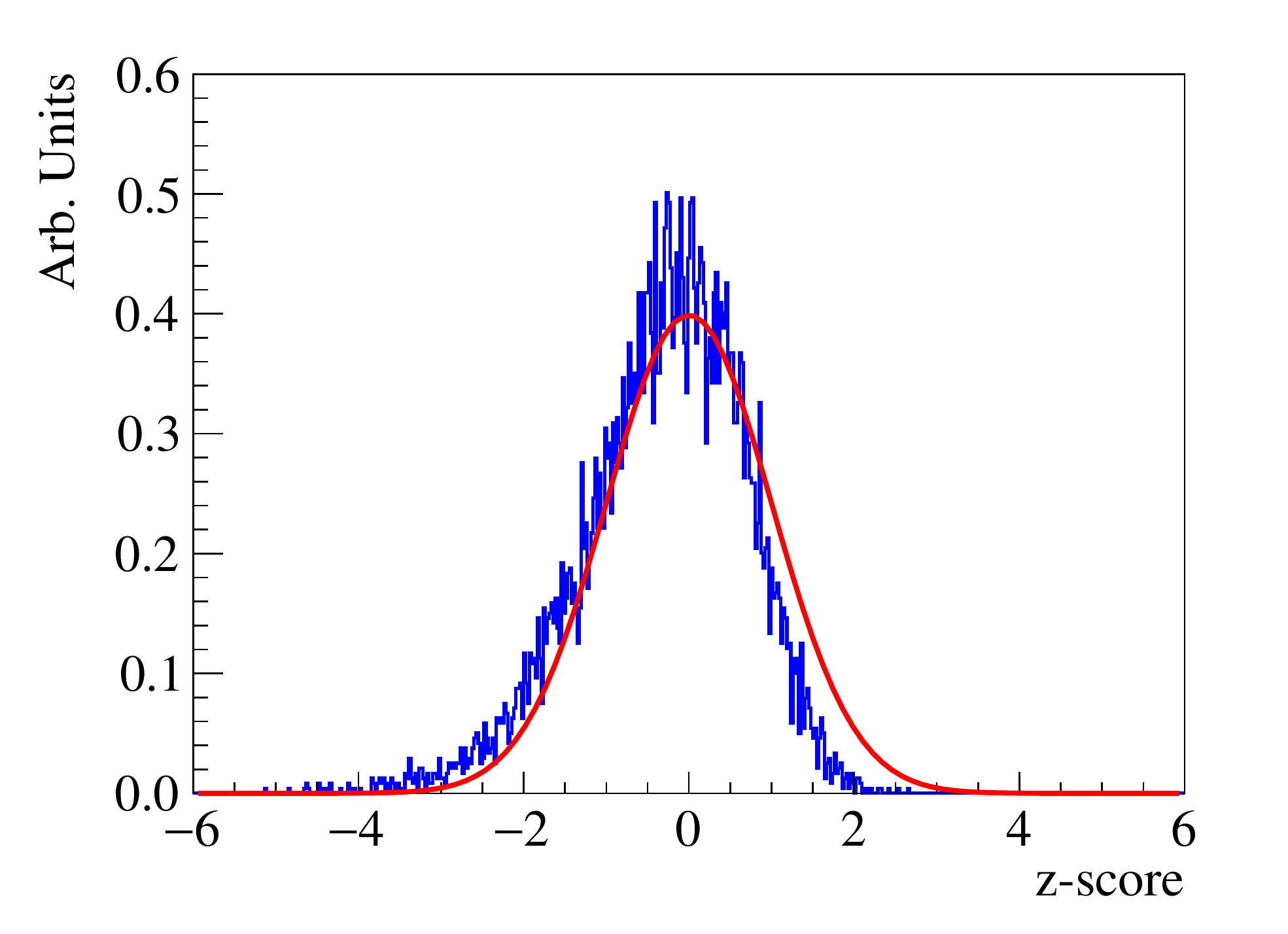}
    \caption{$X_0 = 5~\mathrm{\mu Bq/kg}$, $X_{i} = X_0 \frac{N+1}{2} = 27.5 ~\mathrm{\mu Bq/kg}$}
    \label{fig:zstatN-3X50}
  \end{subfigure}
  \begin{subfigure}[t]{.5\textwidth}
    \includegraphics[width=\textwidth]{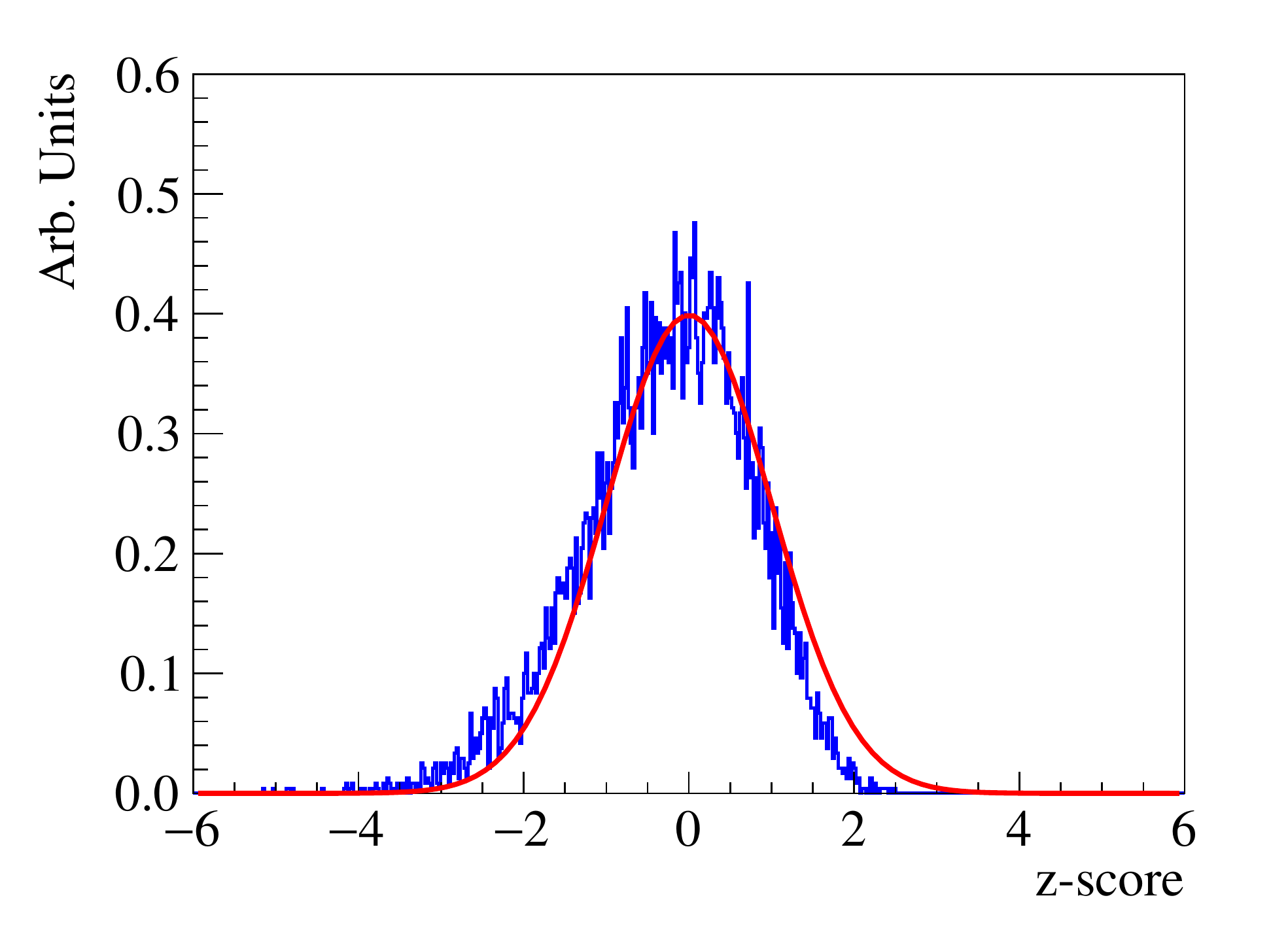}
    \caption{$X_0 = 5~\mathrm{\mu Bq/kg}$, $X_{i} = X_0 (N+1-i)$}
    \label{fig:zstatN-2X50}
  \end{subfigure}
  \end{multicols}
\caption{Distributions of $z$-scores for the realistic scenarios considered in Section \ref{sec:realistic} are plotted in blue
and the standard Gaussian ($\mu=0,\sigma=1$) is drawn in red for comparison.
 (Colors online)
}
\label{fig:zstat}
\end{figure}

Shown in Figure \ref{fig:zstat} are the distributions of the $z$-scores of 10000 instances of assay campaigns
for the three realistic scenarios with $X_0$ set to 10 $\mu$Bq/kg and 5 $\mu$Bq/kg, and
Table \ref{tab:coverage} shows their 1-, 2- and 3-$\sigma$ coverage probabilities.
When the total background is dominated by parts with high $X$ (Scenarios 2 and 3), 
the deviations of the $z$ distributions from Gaussian are small, 
and the coverage probabilities are close to what is expected of a Gaussian distribution. 
In Scenario 1 where low-$X$ parts dominate, we see a significant deviation from the Gaussian.
This is expected
since 
our sensitivity projection method 
generally 
underestimates sensitivity in such cases
as seen in Figures \ref{fig:realisticN-1X100-sens} and \ref{fig:realisticN-1X50-sens},
while $\sigma_S$ only quantifies the statistical uncertainty arising from assay measurements.
The amount of deviation can be reduced by 
improving assay techniques. 

\begin{table}
\centering
\begin{tabular}{c|ccc|ccc||c}
& \multicolumn{3}{c|}{$X_0 = 10$} & \multicolumn{3}{c||}{$X_0 = 5$} & \\
$\sigma$ & Scenario 1 & Scenario 2 & Scenario 3 & Scenario 1 & Scenario 2 & Scenario 3 & Gaussian \\
\hline
1 & 0.685 & 0.692 & 0.693 & 0.303 & 0.708 & 0.684 & 0.683 \\
2 & 0.914 & 0.958 & 0.962 & 0.746 & 0.945 & 0.944 & 0.955 \\
3 & 0.975 & 0.995 & 0.997 & 0.922 & 0.989 & 0.992 & 0.997 \\
\end{tabular}
\caption{Coverage probabilities for the three realistic scenarios with $X_0$ = 10 and 5, 
considered in Section \ref{sec:realistic}. }
\label{tab:coverage}
\end{table}

\section{Conclusion}
\label{sec:Conclusion}

We have created a model of the sensitivity projection for 
a hypothetical 0$\nu\beta\beta$ experiment, 
starting from the assay of detector materials, 
up to the calculation of the projected sensitivity. 

Using this model, we evaluated how the different interpretations of assay results as priors 
in the sensitivity projection procedure affects the projected sensitivity value. 
We evaluated the results using the notion of ``conservativeness'', 
which is understood as our subjective aversion to overestimating the sensitivity.
This can be quantified by the parameter $\alpha\in[-1,1]$ of an asymmetric loss fuction modified from the usual QLF.
Based on the loss function, for each prior, we can define risk $\rho$, 
which is used as the figure of merit for comparing the priors.


{Based on the figure of merit $\rho_0$, 
we have shown that when assay results produce central values and errors well above the sensitivity of the assay technique, 
the Dirac delta prior is recommended, though its advantage over the Gaussian prior is minimal. 
In the case that the impurity concentrations are below the sensitivity of the assay technique and upper limits are reported, 
the Truncated Gaussian, Gaussian-at-zero, and Uniform priors all give similar results. 
The Truncated Gaussian prior has the advantage that it can be motivated by Bayesian arguments, 
although it requires additional information to be reported by the assayers. 
The Dirac delta prior for upper limits is shown to have 
the worst figure of merit under all but the most ``conservative'' assumptions 
(e.g. $\alpha$ $>$ 0.98 in Scenario 1 considered above.)}

To ease the concern for overestimation,
we also suggested calculating $\sigma_S$ which estimates the error in $S$ 
stemming from statistical fluctuations in the assay process.
Experiments may report ``$S\pm\sigma_S$'' as their projected sensitivity,
explicitly stating the uncertainty due to assay measurements.

\appendix

\section{HPGe sensitivity}
\label{sec:hpgesens}

Let $X$ be the impurity concentration of a sample,
assayed by a HPGe detector with a detection sensitivity of $D$,
and a background rate of $B$, for a livetime of $t$.

The measured sample counts and background counts, $c_s$ and $c_b$,
thus follow the distributions:

\begin{equation}
\begin{cases}
c_s \sim Po(\lambda=D X t + Bt) \\
c_b \sim Po(\lambda=B t) \\
\end{cases}
\end{equation}
where $Po(\lambda=\cdot)$ represents the Poisson distribution with mean $\lambda$.

$X$ can be estimated by $\hat{X}$ which is defined by,

\begin{equation}
\hat{X} = \frac{1}{Dt}(c_s - c_b)
\end{equation}
and its uncertainty is,

\begin{equation}
\hat{\sigma} = \frac{1}{Dt}\sqrt{c_s+c_b}
\end{equation}

On average, to observe a signal, we require 
$\mathbb{E}(\hat{X}) > n \mathbb{E}(\hat{\sigma})$, 
where $n$ is the ``number of sigmas'' required for declaring a detection.
This is satisfied when,

\begin{equation}
X > \frac{n^2}{2 D t} (1+\sqrt{1+\frac{8Bt}{n^2}})
\end{equation}

Now, apply the above inequality to the case described in the main text. 
For $n$ = 1.64 (i.e. 90\% C.L.), $D$ = 1 cps/(Bq/kg), $B$ = 10 cpd, and $t$ = 14 d,
$X > 23.8$ $\mu$Bq/kg. 

\section{Mean vs median upper limit as sensitivity}
\label{sec:avgsens}
\begin{figure}[htbp]
\includegraphics[width=\textwidth]{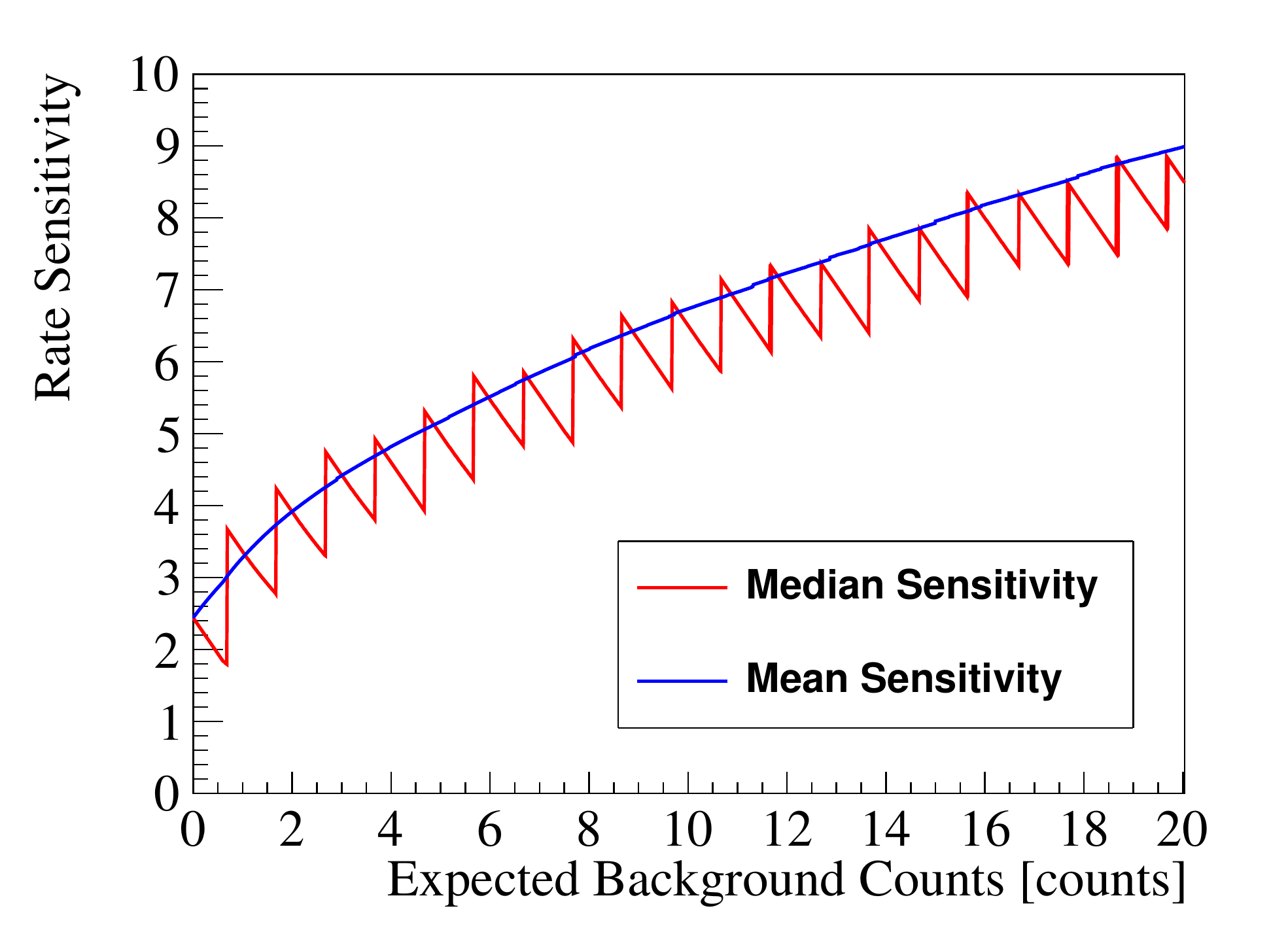}  
\caption{Comparison between using the mean vs the median of upper limits as the sensitivity.
 (Colors online)
}
\label{fig:avgsens}
\end{figure}

One could consider whether to use 
the \textit{mean} or the \textit{median} to calculate an average upper limit. 
If the median is used, the sensitivity as a function of background rate
would exhibit a sawtooth pattern (as shown in Figure \ref{fig:avgsens}). 
This behavior has frequently been observed, 
but this is counter-intuitive as 
it implies that a better sensitivity could be achieved by increasing the background rate
by a small amount just short of reaching the next jump.
However, if the mean is used, the curve becomes monotonous as one would intuitively expect. 
Therefore, in this study, mean sensitivity is used.

\section{Motivating the Truncated Gaussian prior using Bayesian arguments}
\label{sec:bayes}

First, let's clarify the terminology.
As general terminology in Bayesian statistics, 
the (assumed) distribution of a parameter before a measurement is called a \textit{prior} distribution, 
while the distribution of the parameter after incorporating the measurement result is called a \textit{posterior} distribution.
In sensitivity projections, there are two stages of measurements: 
the first stage is the assay measurements and 
the second stage is the actual experiment that attempts to detect a rare physical process.
The priors described in the main text are called as such since they are priors relative to the actual experiment.
Though, at the same time, they are \textit{posteriors} relative to the assay measurements. 
(To avoid confusion, when mentioned in this appendix, the priors in the main text are in quotes from now on.)
In other words, the ``priors'' can, in fact, be derived using Bayes' theorem by assuming prior distributions 
for the impurity concentration $X$ and the HPGe background rate $B$.
Using the same parameter definitions as in \ref{sec:hpgesens},
the posterior distribution for $X$ and $B$ given $c_s$ and $c_b$ can be 
written as follows by applying Bayes' theorem,

\begin{equation}
\begin{aligned}
P(X, B | c_s, c_b) &= A P(c_s, c_b | X, B) P (X, B) \\
& = A P(c_s | X, B) P(c_b | B) P(X) P(B) \\
& = A \frac{\lambda_s^{c_s} e^{-\lambda_s}}{c_s!} \frac{\lambda_b^{c_b} e^{-\lambda_b}}{c_b!} P(X) P(B) \\
\end{aligned}
\end{equation}
where $A = \frac{1}{P(c_s) P(c_b)}$, $\lambda_s = DXt+Bt$, and $\lambda_b = Bt$.

\begin{figure}[htbp]
\includegraphics[width=\textwidth]{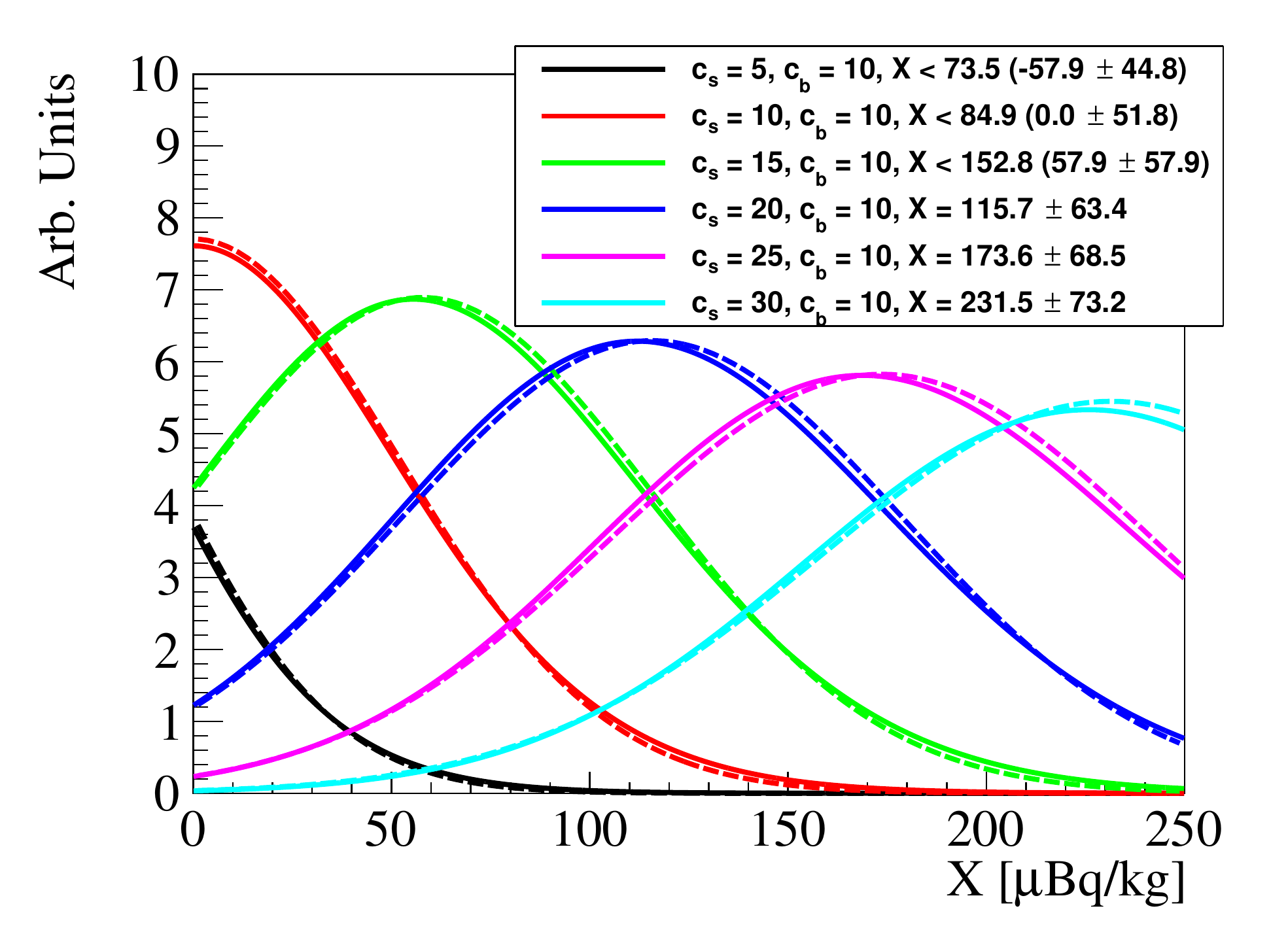}  
\caption{Comparison between posteriors derived using Bayes' theorem assuming uniform priors for X and B (solid lines) 
and Truncated Gaussian ``priors'' (dashed lines) for some values of $c_s$ and $c_b$. 
Here, $D = 1$ and $t = 1$ d. With such parameters, having $X > 102$ $\mu$Bq/kg is required to observe a signal on average.
 (Colors online)
}
\label{fig:bayes}
\end{figure}

Now, assume the priors $P(B)$ and $P(X)$ are (improper) uniform distributions defined in $[0,\infty)$.
Then, marginalizing over $B$, which is a nuisance parameter, gives,

\begin{equation}
\begin{aligned}
P(X | c_s, c_b) &= \int_0^\infty P(X, B | c_s, c_b) dB \\
& = A e^{-x} x^{c_s+c_b+1} U(c_b+1,c_s+c_b+2,2x)\\ 
\end{aligned}
\end{equation}
where $x = D X t$, and
$U(\cdot,\cdot,\cdot)$ is the
confluent hypergeometric function of the second kind.
We can then compare $P(X | c_s, c_b)$ 
with the corresponding Truncated Gaussian ``priors''
for the same $c_s$ and $c_b$.
As shown in Figure \ref{fig:bayes}, 
$P(X | c_s, c_b)$ and its corresponding Truncated Gaussian ``prior'' are very similar in shape.
Therefore, the Truncated Gaussian ``prior'' can be seen as 
an approximation to $P(X | c_s, c_b)$ which has a more complex functional form.

Another feature of $P(X | c_s, c_b)$ (and its Truncated Gaussian approximation) that other ``priors'' lack
is that the transition between the upper limit and the central value regions is smooth. 
This means that the choice of C.L. by the assayer 
to report an upper limit 
has no effect on the ``prior'' shape,
as the upper limit is not directly used in the definition of the ``prior''.


\section*{Acknowledgements}
\label{sec:Acknowledgements}
This work was supported by the Nuclear-physics, Particle-physics, Astrophysics, and Cosmology (NPAC) Initiative, a Laboratory Directed Research and Development (LDRD) effort at Pacific Northwest National Laboratory (PNNL). PNNL is operated by Battelle for the U.S. Department of Energy (DOE) under Contract No. DE-AC05-76RL01830.

This research was done using resources provided by the Open Science Grid (OSG), which is supported by the National Science Foundation and the U.S. Department of Energy's Office of Science.

R.H.M.T. would like to thank the organizers of the OSG User School 2017 in Madison, Wisconsin, for their hospitality and
for offering the opportunity to learn and use the OSG.  
The authors would also like to thank Ben Loer, Kevin Anderson, Paul Eslinger, Scott Cooley, and Jan Strube
for their valuable feedback and suggestions.

\bibliographystyle{elsarticle-num} 
\bibliography{upperlimits.bib}

\nocite{*}

\end{document}